# Agile manoeuvring of dandelion-inspired micro-flyers with vortex-enabled stability

Jianfeng Yang,[1] Soumarup Bhattacharyya,[2] Aditya Potnis,[2] Hao Zeng,[1]* Ignazio Maria Viola[2]*

**Affiliation:**

[1] Faculty of Engineering and Natural Sciences, Tampere University, P.O. Box 541, FI-33101 Tampere, Finland.

[2] School of Engineering, Institute for Energy Systems, University of Edinburgh, Edinburgh, EH9 3FB, UK.

*Correspondence to: hao.zeng@tuni.fi; i.m.viola@ed.ac.uk

**Abstract**.

Manoeuvring untethered, centimetre-scale airborne structures has been a long-standing challenge. Active flight systems, relying on high-power-density actuators alongside mechanical and electronic components, are constrained by critical limitations in energy delivery and miniaturisation. In contrast, passive systems transported and distributed by the wind typically lack the capability for mid-air controlled manoeuvrability. Here we report an ultra-light (1.2 mg) hexagonal polymeric assembly capable of passive flight with optical control of its trajectory. This dandelion-inspired micro-flyer incorporates six radially arranged filamentous structures, of which morphology is dynamically controlled through photomechanical deformation by six independent soft actuators made of liquid crystalline elastomer thin films. Compared to the diaspore of the dandelion (*Taraxacum officinale*), micro-flyer demonstrate a similar terminal velocity (~0.5 m s$^{-1}$), 45% better positional stability and nearly zero rotational rate (1.68 ± 1.0° s$^{-1}$; natural seeds: 50.8 ± 17.7° s$^{-1}$). Particle image velocimetry reveals that a stable asymmetric separated vortex ring underlies its flight stability, enabling mid-air steerability. When free-falling in a low-turbulent airstream, the light-driven hexapodal fliers demonstrate precise altitude control, reversible body flipping, pattern formation, interactive swarm, and controlled trajectories across three-dimensional space. The results show that responsive materials with light-induced asymmetry can bring about manoeuvrability in air, paving the way for agile, untethered controlled micro-fliers.

**Main text.**

Micro and nano drones have undergone a transformative evolution, progressing from rudimentary motor-driven prototypes[1] to advanced piezoelectric[2] and dielectric-actuated systems[3], and ultimately to sophisticated platforms employing novel actuation paradigms[4-7]. This developmental trajectory has been driven by interdisciplinary advancements spanning engineering innovations, insights from biological flight strategies[8,9] and materials sciences[6,10,11]. Among these, active drones that rely on onboard actuators or propulsion units for thrust and manoeuvrability represent a pinnacle of engineering miniaturisation[12,13]. However, they are limited by the need for high power density and the complexities of mechanical transduction[5]. To meet their power demands, these systems often resort to tethered setups that undermine steerability and operational range[4,5,8]. Moreover, achieving compact integration of power units, control electronics, and actuation elements limits the development of agile, untethered drones with autonomous flight capabilities[14,15].

In contrast, nature offers an energy-efficient alternative for centimetre-scale flight: wind-dispersed seeds are passively transported and distributed by the wind and remain airborne without relying on active propulsion[16,17]. For example, maples exploit autorotation to generate lift and decrease the terminal velocity[18,19], while dandelions exploit a highly porous filamentous pappus to form a Separated Vortex Ring (SVR)[20] associated with enhanced drag. The maple's lift and the dandelion's enhanced drag allow these diaspores to fall slowly and be uplifted by turbulent updrafts, remaining afloat for hours and being transported for hundreds of kilometres[21,22]. The intrinsic efficiency of these passive locomotion mechanisms circumvents many of the energy issues that limit active systems, while simplifying design, reducing structural size and weight[6,23-29].

Notwithstanding the above-mentioned advantages, achieving control in passive flight presents critical engineering challenges. The principal limitation resides in enabling untethered morphology control to attain precise manoeuvrability within the Three-Dimensional (3D) space while retaining aerodynamic stability and control of the trajectory. Rapidly deformable, light-responsive materials have been demonstrated to enable take-off and landing actions[26,30], but efforts to extend this manoeuvrability to sustained aerial navigation have been hampered by the instability issue. The relatively low inertia of these devices compared to heavier powered drones often results in erratic motion patterns, including non-stop spinning and large-scale oscillations[26], thereby complicating the maintenance of a stable flight path.

Here, we report a micro-flyer with unprecedented aerodynamic stability and steering capability. Figure 1a shows the hexapodal flier inspired by the radially distributed porosity of the pappus of the dandelion's diaspore. It features six cento-aligned robotic arms made of artificial muscles. Each arm is uniformly tipped with nine biomimetic filaments made of fabric fibre. To achieve untethered light actuation, a Liquid Crystal Elastomer (LCE)[31,32] strip is used as an artificial muscle to induce mechanical deformation. The details of design, photothermal actuation mechanism, and optical properties of LCE are reported in the Supplementary Note 1 (Supplementary Figs. 1-13).

Alike the filamentous pappus of the dandelion diaspore (Supplementary Figs. 13), the hexapodal structure enables the generation of a SVR (Fig. 1b, c) that enhances the aerodynamic drag and diminishes the terminal velocity, $u_\text{t}$ (in the closed state, $u_\text{t}$ = 0.73 ± 0.04 m s$^{-1}$; in the open state, $u_\text{t}$ = 0.44 ± 0.02 m s$^{-1}$; for the natural dandelion diaspore, $u_\text{t}$ = 0.5 ± 0.05 m s$^{-1}$). Details of the SVR are reported in Supplementary Note 2 and Supplementary Figs 14-17. The drag coefficient ($C_\text{D}$) of micro-flyers within a range of sizes (diameter $d$ from 1.5 cm to 3 cm) and weights (from 1.12 mg to 4.02 mg) was evaluated through drop tests in quiescent flow (details in Supplementary Fig. 18 and Methods). The opening angle $\alpha$ between filaments at opposite ends can be controlled by varying the excitation light intensity, yielding to a reduction in $u_t$ by more than 2/3 when filaments are turned from closed (vertical, $\alpha = 0°$) to open (horizontal, $\alpha = 180°$) (Supplementary Fig. 19). Furthermore, the tuneable initial curvature of the LCE allows it to be programmed into either a curved or flat configuration[33]. This design flexibility enables the micro-flyer to adopt two distinct initial structural states: a closed or open configuration. Light-triggered ascent or descent can be achieved through variations in $C_D$ and $u_t$ by morphing the micro-flyer structure (Supplementary Fig. 20).

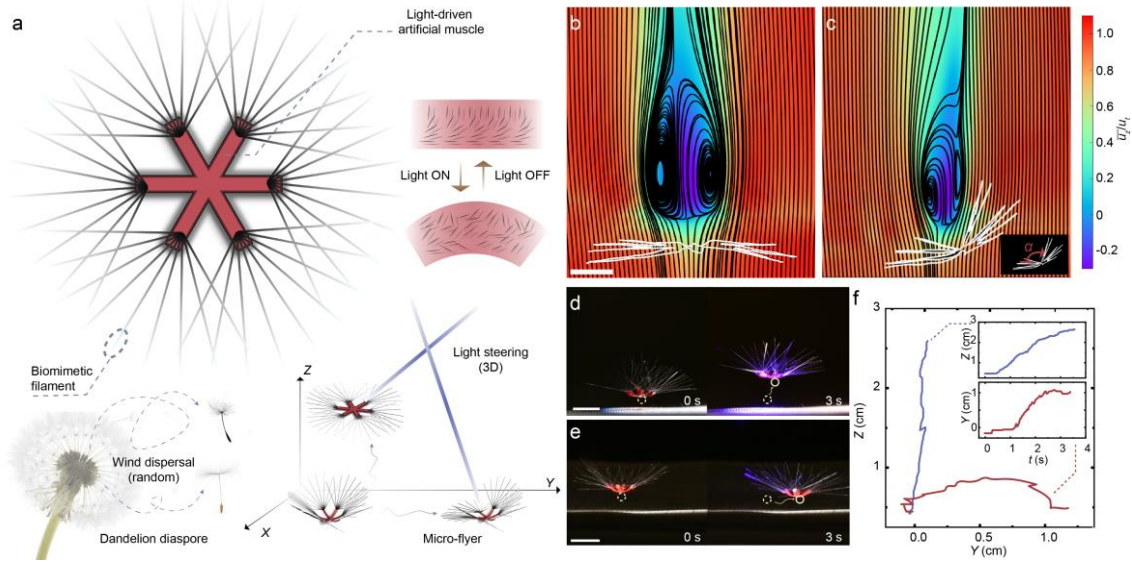

**Fig. 1. System concept.** (a) The schematics of the micro-flyer and of its light-driven artificial muscle when the external illumination is on and off. Ratio of the magnitude of the time-averaged vertical flow velocity $\overline{u_z}$ and of the terminal velocity $u_t$ on a vertical plane through the centre of the micro-flyer for an opening angle (b) $\alpha = 180°$, and (c) $\alpha = 120°$ (asymmetric configuration). Light-induced (d) upward and (e) sideward displacement of the micro-flyer. Wind tunnel speed is (d, e): 0.6 m s$^{-1}$. (f) Upward (blue) and rightward (red) trajectories corresponding to Fig.1d and 1e, respectively, with their time evolution presented in the insets. Scale bars are 0.5 cm. Light intensity is 700 mW cm$^{-2}$.

To investigate the aerodynamics of the micro-flyer, a bespoke vertical wind tunnel (Supplementary Fig. 21) was built to produce a uniform upward airflow matching the terminal velocity, so that the micro-flyer hovers at a constant height in the wind tunnel (Methods). The micro-flyers with any $\alpha$ exhibit remarkable stability as compared to natural seed diaspore (Supplementary Fig. 22). The stability is defined as the standard deviation ($\sigma$) of the horizontal ($Y$) and vertical ($Z$) position over a 1 s time interval measured at 25 Hz. Specifically, while the natural dandelion diaspore exhibits $\sigma = 0.028$ cm ($Z$ axis; $n = 10$ independent biological repeats), the micro-flyer shows $\sigma = 0.016$ cm ($Z$ axis; $n = 5$ repeated measurements of the same sample), *i.e.* 45% less than the natural counterpart (Supplementary Fig. 22). Furthermore, the micro-flyer possesses ultra-high angular stability around its vertical axis compared to the dandelion diaspore. While the spinning rate of the dandelion diaspore is $50.8 \pm 17.7$ ° s$^{-1}$ ($n = 10$), that of the micro-flyer is only $1.68 \pm 1.0$ ° s$^{-1}$ ($n = 5$, Supplementary Fig. 23). The stability of airborne position and a nearly zero spin are the key prerequisites for enabling complex, multi-mode motion control, which will be illustrated later.

Particle Image Velocimetry (PIV) is employed to measure the flow field around the micro-flyer (Methods). For convenience, PIV measurements are taken with the micro-flyer kept fixed at a constant position and with the wind tunnel velocity set at the terminal velocity observed with a free-falling micro-flyer. We demonstrated that the shape and size of the SVR, which is a toroidal vortex in the wake, can be controlled through the opening angle $\alpha$ and by inducing asymmetry in the filamentous structure[34,35] (Figs. 1b, c). Details of aerodynamic measurement are reported in Supplementary Note 3 (Supplementary Figs. 15-23). The SVR size grows with increasing opening angle, as revealed by the increasing distances between both its nodes and the saddle points on a vertical section across its centre (Supplementary Fig. 16). Uniform illumination induces structural opening in an initially closed micro-flyer (Fig. 1d and Supplementary Video 1), resulting in a larger SVR and higher drag, and thus uplifting the micro-flyer in the wind tunnel (displacement data in Fig. 1f). Light induced vertical motion in both configurations are observed in detail for various light intensities (Supplementary Fig. 24-26).

Moreover, impulsively providing a light excitation on one side of the micro-flyer results in the illuminated filament group opening abruptly, generating a lateral force propelling the micro-flyer away from the light, as shown in Fig. 1e, 1f and in the Supplementary Video 2. Details of asymmetric shape-morphing (Supplementary Fig. 27), lateral moving speed (Supplementary Figs. 28 and 30) and light intensity threshold (Supplementary Figs. 29 and 31) reveal the trajectories and the variation in locomotion speed of the micro-flyer at different light intensities.

**Airborne stability.**

For a micro-flyer in steady fall, the weight ($W$) is balanced by the aerodynamic drag ($D$), which increases with increasing velocity (Fig. 2a). To ensure high endurance, the mean terminal velocity must be as low as possible. Furthermore, the stability of the wake, and of the SVR in particular, ensures that the micro-flyer falls steadily, as opposed to chaotically, flattering, tumbling, *etc*. A steady fall is a critical requirement for precise steerability. To investigate the stability of the SVR, contours of instantaneous vertical velocity and streamlines on a vertical plane through the micro-flyer's centre are taken at different time instants (Fig. 2b-d). The small differences in the flow fields at different times confirm that the SVR is stable, similar to that of the dandelion diaspore. The SVR stability persists regardless of variations in the opening angle for both symmetrically oriented models (Supplementary Fig. 15) and asymmetric configurations[36] (Figs. 2e-h).

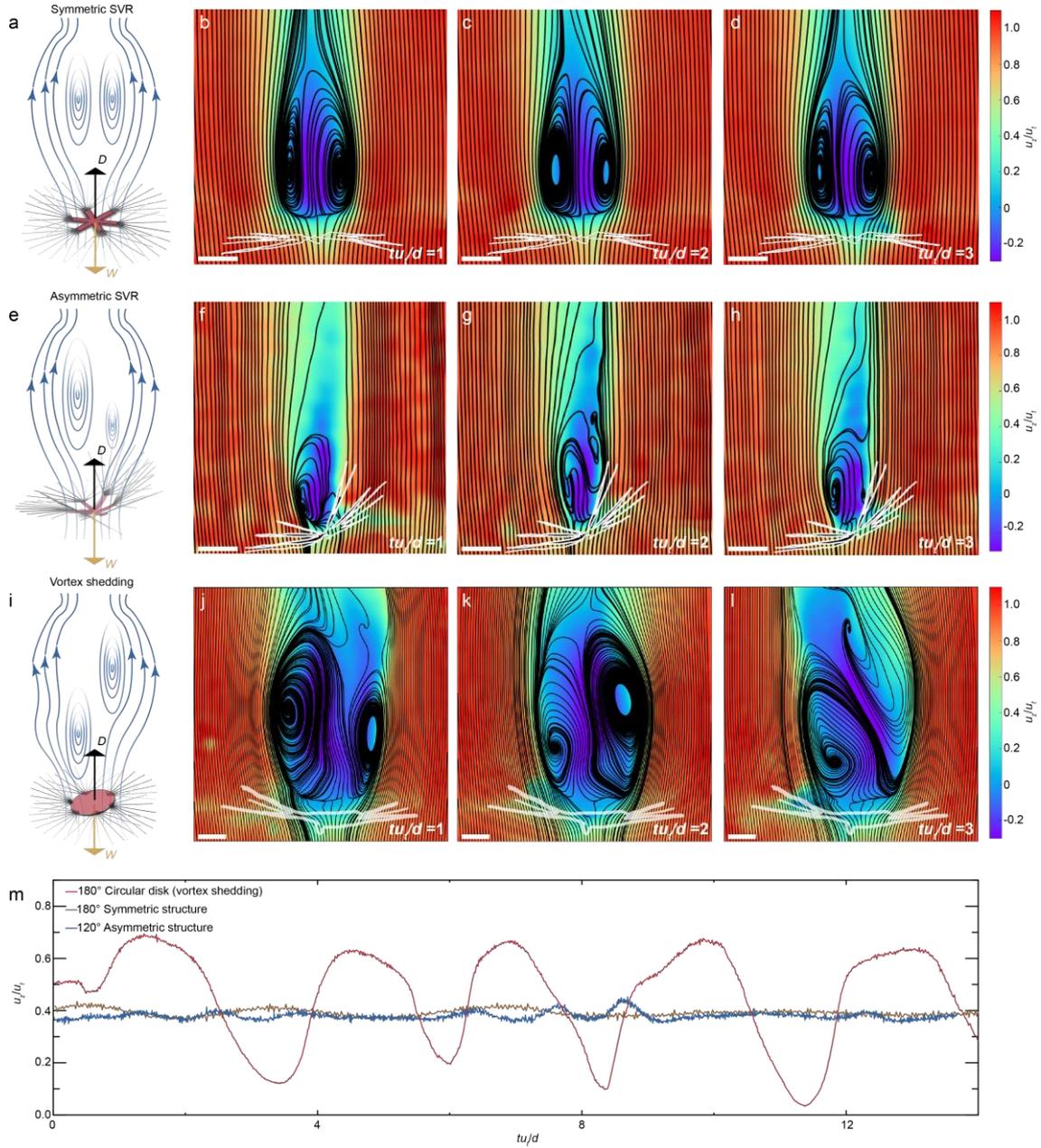

**Fig. 2. Stability of the SVR.** Schematics and instantaneous flow fields of (a-d) the symmetric micro-flyer configuration with an opening angle $\alpha = 180°$ (fully open configuration), (e-h) the asymmetric configuration with $\alpha = 120°$, and (i-l) the fully open micro-flyer ($\alpha = 180°$) with the addition of a 1 cm diameter disk at its centre. The instantaneous flow fields show the contours of the ratio of the magnitude of the instantaneous vertical flow velocity ($u_z$) and of the terminal velocity, and the streamlines constrained on the vertical plane through the centre of the micro-flyer for three instants, each at one convective period apart ($tu_t/d = 1, 2, 3$). (m) Time series of the velocity magnitude one diameter downstream of the centre of the model for the three configurations. Scale bars are 0.5 cm.

As a control, Fig. 2j–l displays contours of instantaneous flow speed and streamlines for a fully open symmetric micro-flyer incorporating an impervious, one-centimetre-diameter disk at its centre[37,38] (Fig. 2i). Unlike the consistent wake topology observed in the micro-flyers, the disk's wake is unstable, resulting in periodic vortex shedding under the same velocity conditions (Supplementary Video 3). This is evident by the low amplitude of the time series of the vertical velocity measured one diameter downstream of the micro-flyer with two different opening angles compared to the large amplitude oscillations of the vertical velocity in the wake of the micro-flyer with the added disk (Fig. 2m). Quantitative analyses of vorticity and detailed comparison between symmetric and asymmetric micro-flyers, as well as hexapodal cores without filaments and impervious disks, are given in Supplementary Fig. 17 and Supplementary Video 3. These findings underscore the pivotal influence of the permeability and structural asymmetry on the vortex dynamics.

**Agile manoeuvring.**

Here we demonstrate the multi-modal manoeuvring of micro-flyers in 3D space. First, the control of the vertical position of a symmetric micro-flyer free flying in a uniform constant upward flow stream is demonstrated, facilitated by the photomechanical change in the opening angle. The altitude $Z$ can be controlled by actively switching on and off the light to symmetrically increase and decrease the opening angle, respectively (Fig. 1f). However, if a constant illumination is provided below the target altitude ($Z_0$) and obscured above that height, the micro-flyer would automatically settle at the target altitude $Z_0$. This procedure, which is demonstrated in Supplementary Fig. 32, allows positioning the micro-flyer at a certain height without actively control or track its position. In fact, consider an initially closed micro-flyer in the illuminated region ($Z < Z_0$). It opens because the provided illumination causes the opening angle and the drag to increase, and thus it moves upwards. When it passes the target height ($Z > Z_0$), it is out of the light beam, and thus the structure closes, reducing the drag and making the micro-flyer to move downwards towards the target position. Overall, the micro-flyer moves upwards when below $Z_0$ and downwards when above it, resulting in a vertically oscillatory kinematics near the desired altitude (Supplementary Figs. 33-35).

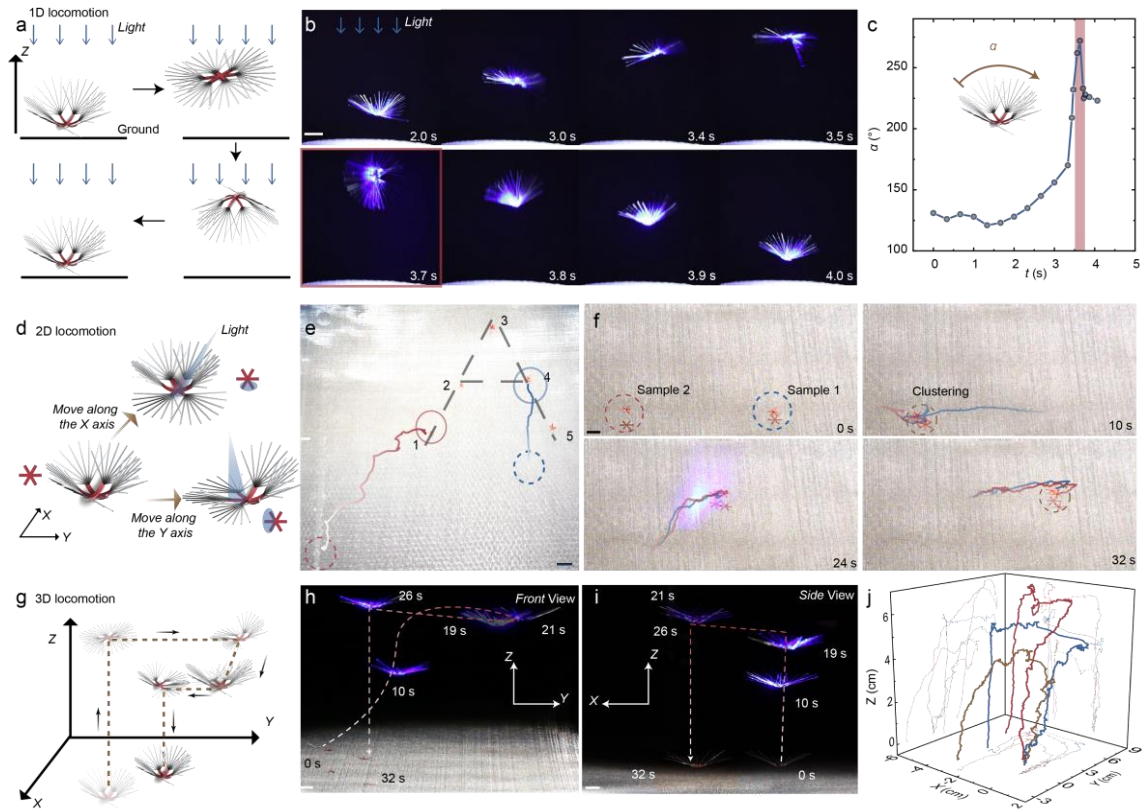

**Fig. 3. Agile manoeuvring of micro-flyer.** (a) Schematic of the light-controlled flipping motion of the micro-flyer. (b) The snapshots of the flipping motion of the micro-flyer. (c) Time evolution of the opening angle α during the flipping process. (d) Schematic of the light-controlled two-dimensional locomotion in the horizontal X-Y plane. (e) Image of five micro-flyers steered on the X-Y plane to reach the vertices of the letter "A". (f) Snapshots of micro-flyer pairs controlled by light. (g) Schematic of the 3D trajectory of micro-flyer. Superimposed images of the micro-flyer performing light-controlled 3D locomotion from (h) the front and (i) side view. (j) 3D trajectory of the micro-flyer with projections onto the *X-Y, X-Z, Y-Z* planes. Wind tunnel speed: 0.6 m s$^{-1}$. Light intensity: 700 mW cm$^{-2}$. The scale bar: 0.5 cm.

A high excitation light intensity causes the opening angle to exceed 180°. The resulting equilibrium position is unstable because the centre of gravity is above the aerodynamic centre, and thus the micro-flyer undergoes a spontaneous flipping to bring the centre of gravity below the filaments (Fig. 3a and 3b shows the schematic drawing and snapshots of the flipping motion, respectively - more details are shown in Supplementary Video 4, and the tracking of trajectory in Supplementary Fig. 36). The shape-morphing induced instability can be qualitatively assessed by tracking the opening angle $\alpha$ (inset of Fig. 3c). The micro-flyer flips as $\alpha$ reaches 260° at 3.5s. Thereafter, $\alpha$ stabilises at approximately 220°; the light now illuminates the opposite side of the micro-flyer. (Fig. 3c). After ceasing the light, the micro-flyer flips back and returns to its original configuration. The flipping is predictable and repeatable, as demonstrated by a series of eleven successive cycles in Supplementary Fig. 37.

While the above one-dimensional (1D) manoeuvres concern only the vertical direction, two-dimensional (2D) manoeuvres on the horizontal *X-Y* plane are presented in the following. Uneven light excitations between arm segments introduce an asymmetric deformation. Specifically, selective illumination of the left or right tri-legged segments results in translational motion towards right or left, while targeting the front or rear segments enables movement backwards or forwards (Fig. 3d). This allows targeting any position in the horizontal plane. The fine-tuning of the 2D positioning among five individual micro-flyers to form the shape of the letter "A" is demonstrated in Fig. 3e and Supplementary Video 5. Airborne patterns forming the letters "A", "B" and "C" are reported in Supplementary Figs. 38-40.

Mutual attraction was observed when two micro-flyers are at a distance lower than 3 cm, which is equivalent to one micro-flyer's diameter. After interacting, two micro-flyers form a cohesive unit that can subsequently be controlled like a single unit (Supplementary Fig. 41, Fig.

3f and Supplementary Video 6). This cluster exhibits approximately the same terminal velocity and stability as an individual unit. The trajectory can be controlled by manipulating just one micro-flyer inside the cluster, thereby enabling synchronised motion of the swarm. The 2D trajectory control of swarms of three micro-flyers is reported in the Supplementary Fig. 42.

Finally, complex 3D trajectories are demonstrated, where the micro-flyer is steered in different horizontal directions by choosing which of the six robotics arms is actioned (Fig. 3g). By manual handling of an optical fibre head inside the wind tunnel, a micro-flyer is steered to ascend to a predefined height, perform a smooth lateral translation and subsequently descend near its original position (Fig. 3h, 3i and Supplementary Video 7). The 3D steering with trajectory control across three distinct target altitudes is shown by the representative paths in Fig. 3j (Supplementary Figs. 43, 44).

**Discussion and conclusion.**

Although we have achieved an agile manoeuvring over 3D space (Fig. 3h-j), asymmetric deformation of the micro-flyer inevitably results in vertical drift (Supplementary Figs. 45, 46). This is because asymmetric deformation alters the drag, leading to either an upward or downward movement. This reduces the capacity for 3D steering (Fig. 3g–3i) along a precise trajectory.

Integrating stimulus-response material with passive flight mechanisms represents an effective approach to control the flight of insect-scale flyers that are passively transported and distributed by the wind. In this study, light stimulation has been selected as the primary actuation method due to its well-established applicability. However, this approach can be extended to any external stimulus (*e.g.* humidity, heat) capable of promoting the change in morphology[39,40], potentially unlocking a wide range of capabilities and operations. Furthermore, fully autonomous tasks would be possible if a sufficiently light energy source could be carried without excessively increasing the terminal velocity, limiting the endurance. The micro-flyers equipped with sensing and communication capabilities have already been demonstrated[23], and current efforts are exploring biodegradable and self-degrading variants to eliminate the need for post-mission retrieval. The precise manoeuvrability shown in this study expands the potential applications of these systems, including environmental monitoring, surveillance, and target recognition. One particularly promising area is the monitoring of climate change. The 2021 Global Climate Observing System Status Report[41] identified gaps in existing observing systems that hinder an in-depth understanding of climate change and the identification of effective mitigation actions. The report highlights the critical need for increasing sensitivity, resolution, and coverage of in situ measurements. While conventional sensing infrastructure remains spatially constrained and autonomous drones offer limited operational persistence, the micro-flyers—when outfitted with onboard sensors and deployed in swarms—could provide scalable, high-resolution, and long-duration in-situ monitoring. Their silent operation, low ecological impact, and ease of deployment make them especially well-suited for sensitive or remote environments.

Furthermore, their demonstrated steerability enables complex, coordinated aerial behaviours, offering a powerful new platform for distributed sensing systems. While further development is needed for large-scale deployment, this work lays the foundation for soft, untethered, and environmentally adaptive microflyers that could transform how we observe and interact with the natural world.

To conclude, drawing inspiration from the aerodynamics of dandelion diaspores, we developed ultralight hexapodal micro-flyers that harness photomechanical liquid crystal elastomers to achieve unprecedented control and manoeuvrability. They have a weight of the order of one milligram, similar to their natural counterpart, and a range of terminal velocities from 88% to 146% of the mean terminal velocity of the dandelion diaspore. The shape-morphing of the hexapodal structure creates asymmetric, separated vortex rings and a lateral force, enabling the control of their horizontal displacement. In wind tunnel experiments, micro-flyers demonstrated superior linear and rotational stability compared to the natural dandelion. The analysis of the flow field reveals high aerodynamic stability in every operational configuration. We have demonstrated the control of the height inside an upward free airflow, light-induced body flipping, formation of two-dimensional patterns, swarming and navigation through complex three-dimensional trajectories. This study significantly simplifies structural design, reduces weight, and enhances energy efficiency, expanding the potential applications of micro-flyers for long-range and long-duration sensing missions.

## Methods

**Material.** 1,4-Bis-[4-(6-acryloyloxyhexyloxy) benzoyloxy]-2-methylbenzene (99%, RM82) was obtained from SYNTHON Chemicals. 6-Amino-1-hexanol and dodecylamine from TCI, 2,2-Dimethoxy-2-phenylacetophenone were obtained from Sigma-Aldrich, Disperse Red 1, and Disperse Blue 14 were obtained from Merck. All chemicals were used as received.

**Fabrication of LCE film.** Liquid crystal cells were prepared by assembling two coated glass substrates. One substrate was coated with polyvinyl alcohol (PVA, 5 wt% in water, spin-coated at 3000 RPM for 1 min, and baked at 90°C for 10 min) for uniaxial alignment, while the other was coated with polyimide (PI, spin-coated at 3000 RPM for 1 min, and baked at 180°C for 20 min) to achieve homeotropic alignment. A 5 μm microsphere spacer (Thermo Scientific) was used to control the gap between the glass slides, determining the thickness of the LCE film. The liquid crystal mixture, consisting of 0.3 mmol RM82, 0.115 mmol 6-amino-1-octanol, 0.115 mmol dodecylamine, and 2.5 wt% Irgacure 651, was melted at 85°C and infiltrated into the cell via capillary action. The cell was held at 85°C for 10 min, then cooled to 63°C at a rate of 1°C min$^{-1}$ and stored at 63°C for 24 h to facilitate the aza-Michael addition reaction (oligomerization). Polymerisation was completed by UV irradiation (365 nm, 180 mW cm$^{-2}$ for 10 min). After opening the cells with a razor blade, 1 mg of Disperse Red 1 was spread on the sample surface and diffused into the elastomer on a hot plate at 100°C for 10 min.

**Sample assembly.** Three LCE strips were bonded together to form a hexapodal structure. Each pappus was composed of nine filaments adhered together. The dimensions of the LCE strips were 4 mm × 2 mm × 0.05 mm, with the combined mass of the 54 filaments being 0.66 mg, while the LCE strips themselves weighed 0.54 mg. The total mass of the micro-flyer is 1.20 mg.

**Stimuli sources.** A collimated beam from a continuous-wave solid-state laser (ROITHNER, 2 W, 457 nm) was directed onto the LCE to induce deformation. The absorbed optical power was calculated as the product of the light intensity ($I$) and the illuminated area.

**Drop test.** Drop tests were conducted by releasing micro-flyers of varying sizes from a height of one meter, repeating each trial ten times. The descent was recorded using a digital single-lens reflex camera (Canon EOS 60D) operating at 25 frames per second. Trajectories were analysed using the TRACKER software to determine the terminal velocity ($u_t$), defined as the asymptotic speed at which weight and aerodynamic drag reach equilibrium. The mean terminal velocity between ten repeats was reported. To further investigate the influence of mass on the terminal velocity, we attached calibrated plasticine weights to the micro-flyers and repeated the measurements under identical conditions.

**Vertical wind tunnel.** The tunnel is made of four main sections (Supplementary Fig. 21a): the test section, a section housing the honeycomb, the four-fan array, and the settling chamber. All sections of the tunnel are separated by stainless steel woven wire meshes. The mesh immediately upstream of the test section has a 1.98 mm hole size and 0.559 mm thick wires.

The honeycomb is made of aluminium alloy with a 76.9 kg m$^{-3}$ density and 1/2" cells to suppress horizontal velocity components. Each of the four fans, which are controlled by an Arduino Uno, have a diameter of 0.12 m and a maximum angular velocity of 1050 RPM. Supplementary Fig. 21b presents a photograph of the laboratory setup. In the test section, the mean turbulence intensity $T_u = u'/u_t$ is lower than 2%, where $u'$ is the root-mean-square of the vertical velocity fluctuations.

**Flow visualisation.** Visualisation of the flow as seen in Supplementary Fig. 14 was obtained by overlaying raw PIV images acquired by a high-speed camera. Fifty instantaneous images were stacked for each case, using the ImageJ software, and their average was calculated. The images were further adjusted for brightness and contrast.

**Particle Image Velocimetry (PIV).** DEHS particles were spread from the bottom of the wind tunnel and a 1.5 W continuous-wavelength laser was used to illuminate the DEHS particles on a vertical plane of the test section. High-resolution images were captured with a Fastcam Photron camera equipped with a Tamron 180 mm F3.5 SP AF Di Macro Lens. High-speed image acquisition was performed at frame rates ranging from 500 fps to 1125 fps. The LaVision Davis 10 software was used for the processing of the images and the computation of velocity field, while MATLAB was used for subsequent postprocessing. Pre-processing included subtracting the minimum pixel intensity over 15 images to minimize noise. A multi-pass linear window deformation technique was applied, starting with a 64×64-pixel interrogation window and progressively refining to a 32×32-pixel window in subsequent passes. Sub-pixel displacements were determined using two-dimensional Gaussian regression, with a 75% window overlap. The spatial resolution achieved was 1.55 mm, corresponding to a maximum displacement of 16 pixels at a flow velocity of 0.8 m s$^{-1}$. The particle size averaged approximately three pixels, and the seeding density was 0.025 particles per pixel. To prevent spurious vector generation near the model, geometric masking (polygonal type) was applied in Davis 10, followed by vector interpolation within masked regions. Vectors exceeding a standard deviation threshold of two were replaced to ensure accuracy.

**Impact of the opening angle on the size of the separated vortex region (SVR).** To characterise the SVR, an axis ($X'$) through nodes and one ($Z'$) through the saddle points were introduced. Supplementary Fig. 15e schematically represents the wake structure, with points $N_1$ and $N_2$ marking nodes along line $X'$ and points $S_1$ and $S_2$ marking saddle points along line $Z'$.

## Data availability

The raw data generated in this study have been deposited in Fairdata IDA online storage space at https://doi.org/xxxx. Data are also available from the corresponding author on request.

## Acknowledgments

H. Z. gratefully acknowledges funding from the Academy of Finland (Research Fellowship, No. 340263, and the Flagship Programme PREIN, No. 320165) and support of the European Research Council (Starting Grant project ONLINE, No. 101076207). I.M.V, S.B and A.P acknowledge funding from the European Research Council through the Consolidator Grant 2020 "Dandidrone: A Dandelion Inspired Drone for Swarm Sensing" (H2020 ERC-2020-COG 101001499).

**Author contributions**

I. M. V., H. Z. conceived and supervised the project; J. Y. designed and fabricated the robot. J. Y. performed experiments with helps from S. B. and A. P.; S.B. performed PIV. J. Y., S. B., H. Z. and I. M. V. wrote the manuscript. All authors discussed and contributed to the project.

**Competing interests:** The authors declare no competing interests.

# Supplementary Materials for

# Agile manoeuvring of dandelion-inspired micro-flyers with vortex-enabled stability

Jianfeng Yang,[1] Soumarup Bhattacharyya,[2] Aditya Potnis,[2] Hao Zeng,[1]* and

Ignazio Maria Viola[2]*

**Affiliation:**

[1] Faculty of Engineering and Natural Sciences, Tampere University, P.O. Box 541, FI-33101 Tampere, Finland.

[2] School of Engineering, Institute for Energy Systems, University of Edinburgh, Edinburgh, EH9 3FB, UK.

*Correspondence to: hao.zeng@tuni.fi; i.m.viola@ed.ac.uk

**This PDF file includes:**

1. **Supplementary Notes 1-4.**
   Supplementary Note 1: Design, Mechanism, and Optical Properties of micro-flyer.

   Supplementary Note 2: Stability of the Separated Vortex Ring.

   Supplementary Note 3: Aerodynamic Properties of the micro-flyer.

   Supplementary Note 4: The Multi-Dimensional Manoeuvring of micro-flyer.

2. **Captions for Supplementary Videos 1-7.**

**Supplementary Notes**

**Supplementary Note 1: Design, Mechanism, and Optical Properties of micro-flyers.**

The dandelion-inspired micro-flyer is constructed by integrating liquid crystalline elastomer (LCE) strips with biomimetic filaments (Supplementary Figs. 1, 2). These biomimetic filaments closely resemble natural dandelion filaments in both stiffness (Supplementary Figs. 3, 4) and radius (Extended Data Fig. 13g, h), ensuring comparable aerodynamic performance. The LCE serves as the central active component, enabling mechanical actuation through a thermally induced nematic-to-isotropic phase transition (Supplementary Fig. 5, 6). To achieve light responsiveness, Disperse Red 1 (DR1) is incorporated into the LCE matrix, allowing precise curvature control by modulating illumination intensity (Supplementary Fig. 7, 8). The generated bending forces, ranging from 0.5 mN to 3 mN, are sufficient to support the weight of the filaments (0.00098 mN) (Supplementary Fig. 9) and the aerodynamic force. Additionally, photomechanical durability tests demonstrate stable performance over 100 actuation cycles, with no observable material degradation within the experimental timeframe (Supplementary Fig. 10). All these establish a foundation for photothermally induced shape-morphing of micro-flyer (Supplementary Fig. 11), with wind tunnel experiments confirming that photothermal actuation remains largely unaffected by the wind speed, ensuring robust operation in dynamic airflow conditions (Supplementary Fig. 12).

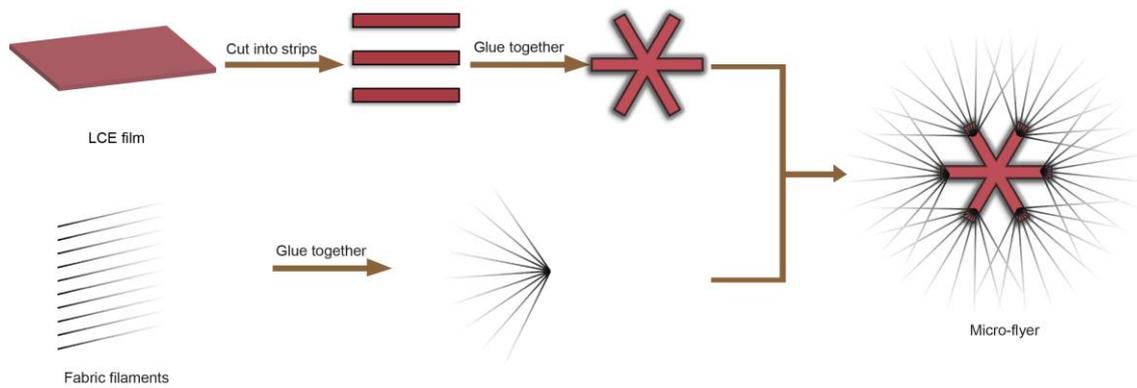

**Supplementary Figure 1. The fabrication process of the micro-flyer.** Schematic illustration of the fabrication process of the micro-flyer.

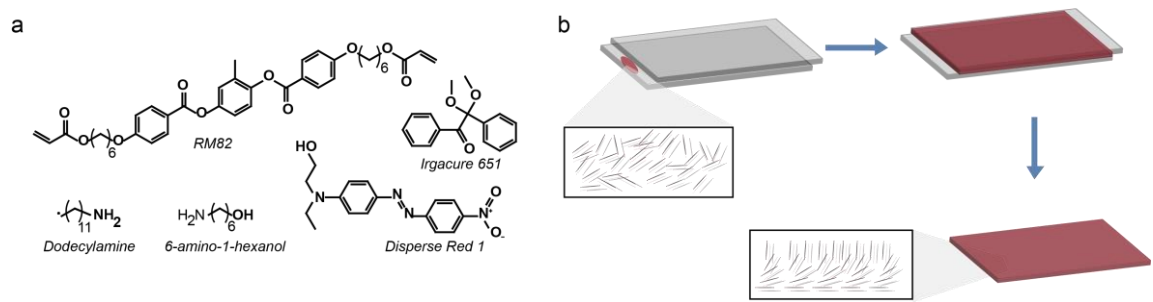

**Supplementary Figure 2. Synthetic steps of the soft actuator.** (a) Chemical structures of all molecules in use. Pre-cured mixture contains 0.3 mmol RM82, 0.115 mmol 6-amino-1-octanol, 0.115 mmol dodecylamine and 2.5 wt% Irgacure 651. (b) Schematic drawing of the LCE film preparation process and molecular alignment.

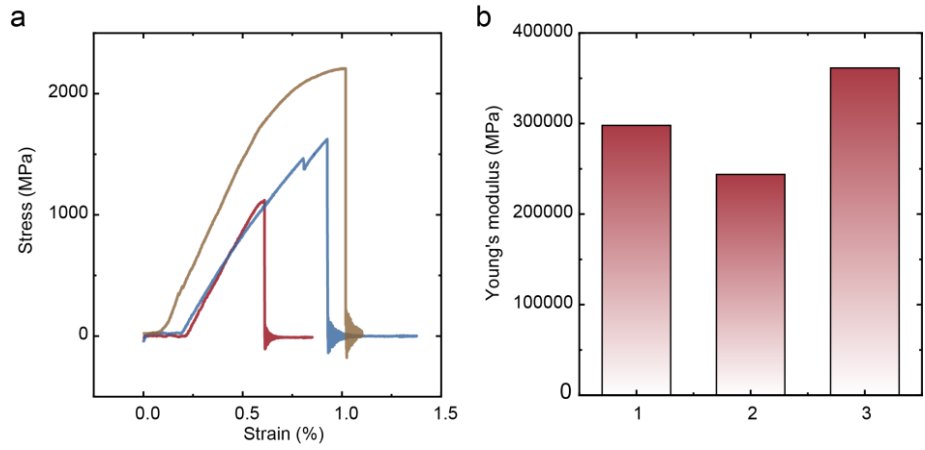

**Supplementary Figure 3. Mechanical properties of biomimetic filaments.** (a) Stress-strain diagram and (b) Young's modulus of fabric filaments for three different samples.

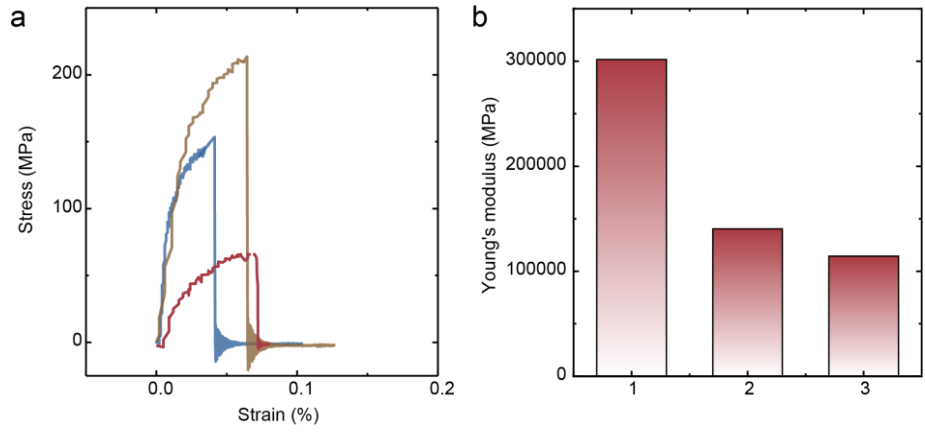

**Supplementary Figure 4. Mechanical properties of natural dandelion filaments.** (a) Stress-strain diagram and (b) Young's modulus of dandelion filaments for three different samples.

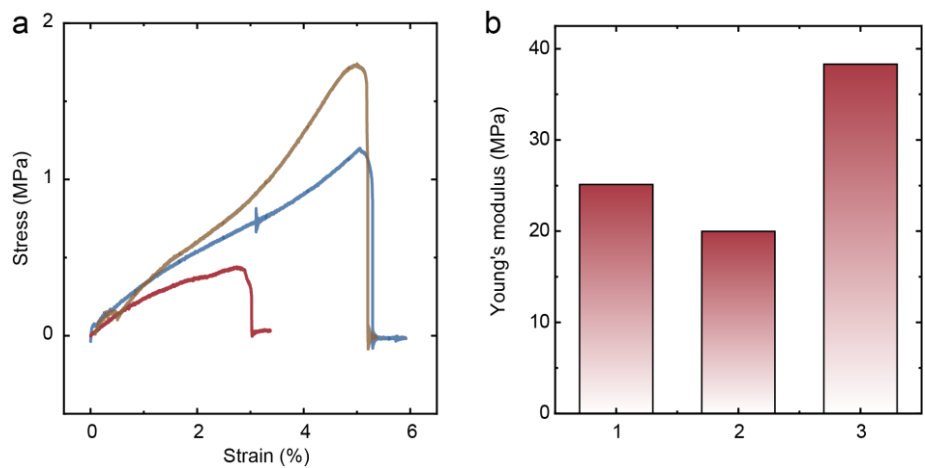

**Supplementary Figure 5. Mechanical properties of the LCE film.** (a) Stress-strain diagram and (b) Young's modulus of LCE strips for three different samples.

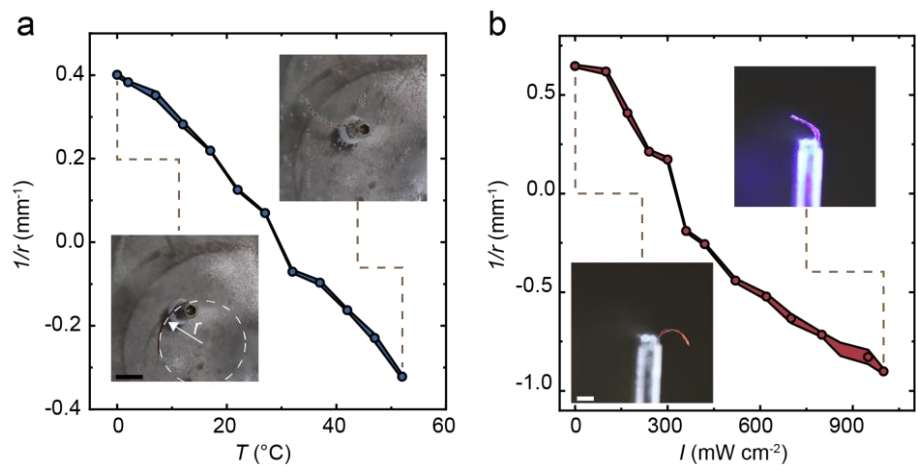

**Supplementary Figure 6. Thermally and optically induced deformations in LCE film.** (a) Curvature (1/*r*, with *r* the radius of curvature) of the LCE strip at different temperatures. Inset: images of the LCE strip at various temperatures. The sample was heated in a temperature-controlled water bath. (b) Change in curvature of the LCE strip under different light intensities. Inset: images of the LCE strip exposed to varying light intensities. All scale bars are 5 mm. The error bars are displayed as mean values +/- standard deviation (n = 3). The same sample was measured repeatedly.

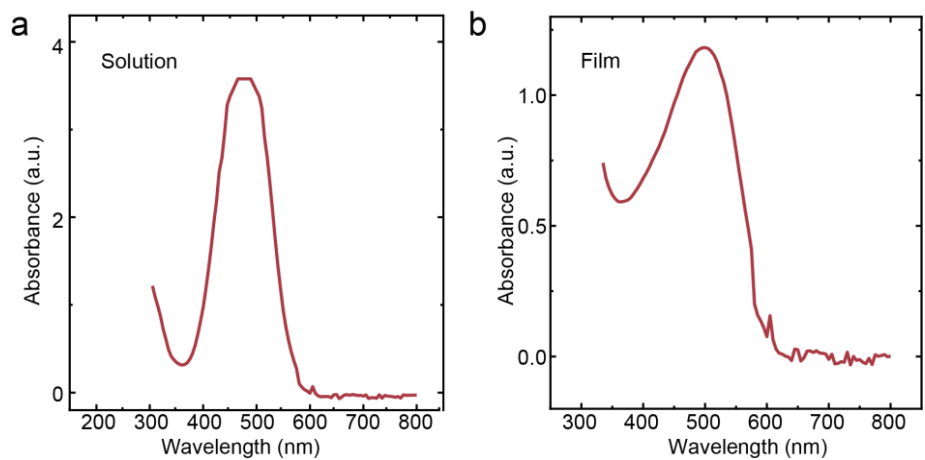

**Supplementary Figure 7. Optical property of the LCE film.** Absorption spectrum of (a) the dye solution and (b) the polymer film after dyeing. Film thickness: 5 μm.

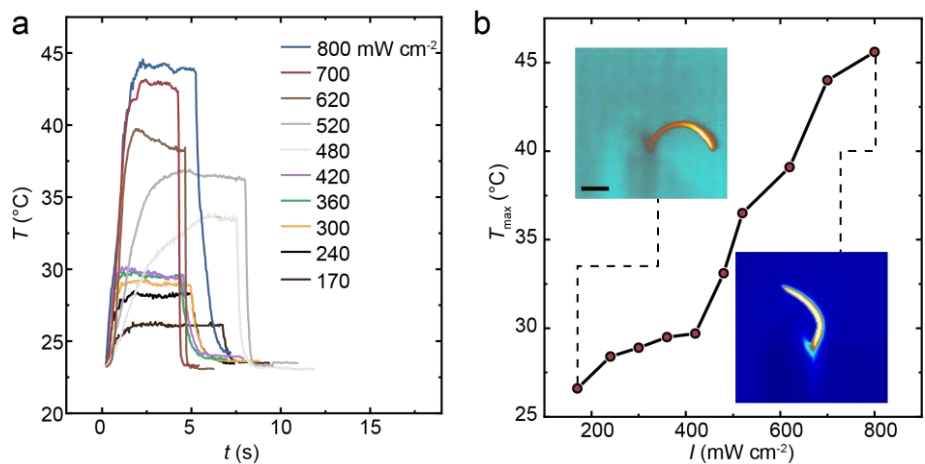

**Supplementary Figure 8. The photothermal kinetics of the LCE film.** (a) The time-history of the temperature during the bending of the LCE film under different light intensities. (b) The maximum temperature ($T_{max}$) achieved under various light intensities. Inset: Infrared image of the LCE film at different light intensities. The scale bar is 5 mm. The dimensions of the LCE film: 15 mm × 3 mm × 0.05 mm.

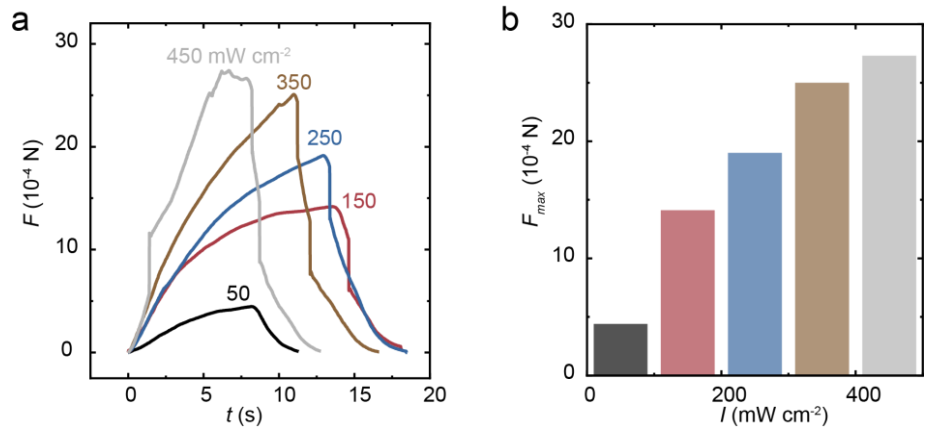

**Supplementary Figure 9. Light-activated forces of the LCE strip.** (a) Time history of the blocking force of a straight LCE strip under different light illuminations. (b) The maximum blocking force of an LCE film upon different light intensities. Strip size for material characterization: $0.7 \times 0.2 \times 0.005$ cm$^3$.

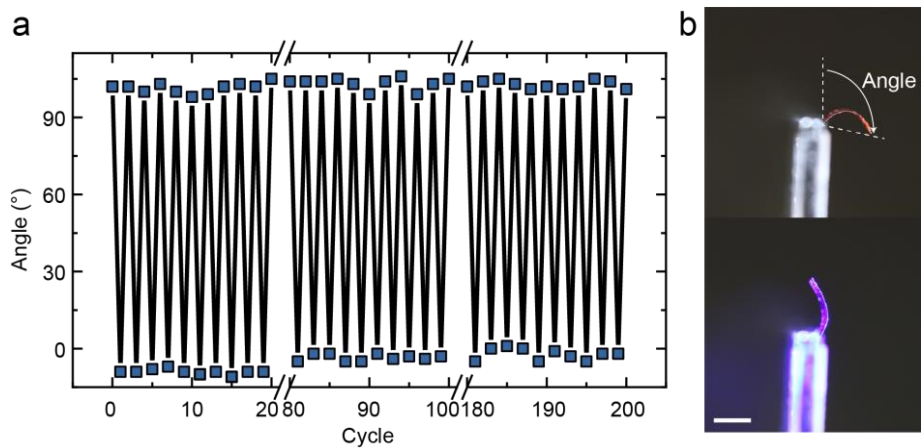

**Supplementary Figure 10. Cycle test of an LCE strip.** (a) Bending angle of an LCE strip during one hundred light actuation cycles. Light: 460 nm, 400 mW cm$^{-2}$. (b) Photographs of light-induced deformation of an LCE strip. The scale bar is 5 mm.

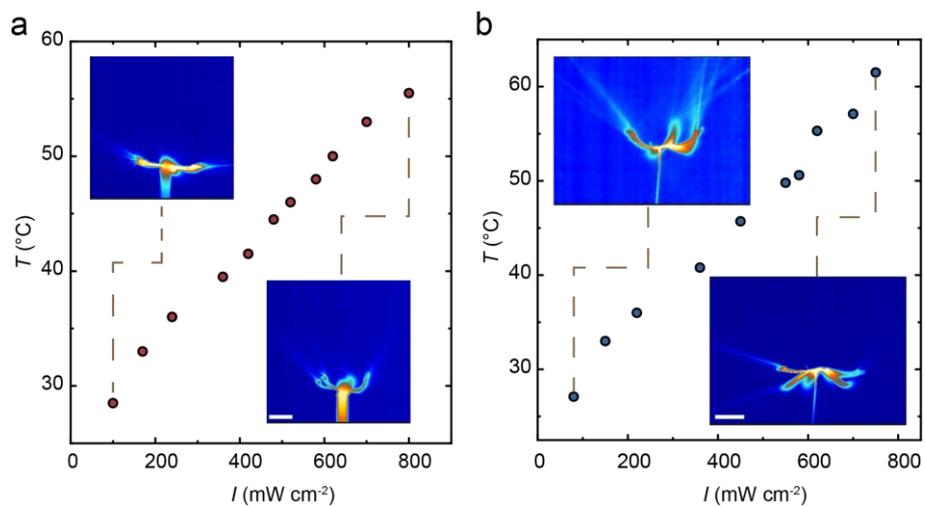

**Supplementary Figure 11. The photothermal property of micro-flyers.** The temperature of the micro-flyer with an initially (a) open and (b) closed configuration upon different light intensities. Insets: Infrared photographs of the micro-flyer with an initially (a) open and (b) closed configuration for two different light intensities. All scale bars: 5 mm.

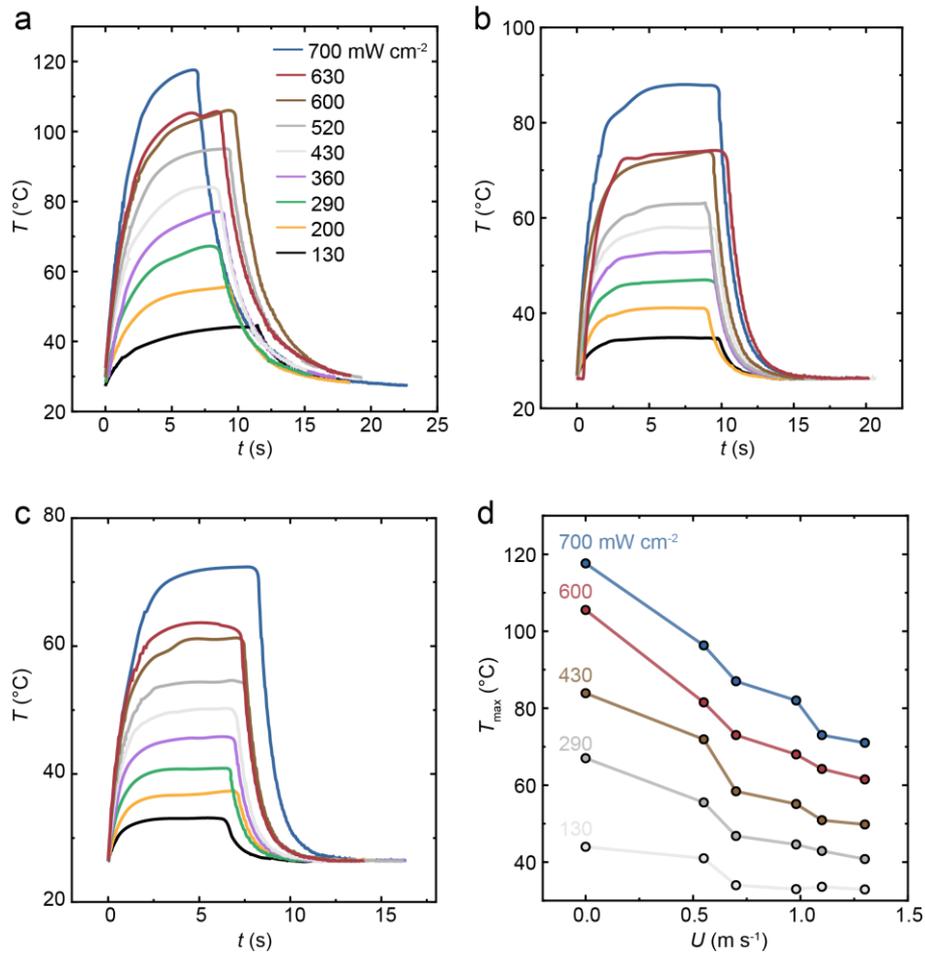

**Supplementary Figure 12. The photothermal property of micro-flyers in different airflows.** Time-history of the temperature of the LCE film bending under varying light intensities in an airflow speed of (a) 0 m s$^{-1}$, (b) 0.7 m s$^{-1}$, and (c) 1.3 m s$^{-1}$. (d) The maximum elevated temperature $T_{max}$ of the LCE film at different flow speeds under different light intensities. The dimensions of the LCE film: 5 mm × 2 mm × 0.05 mm.

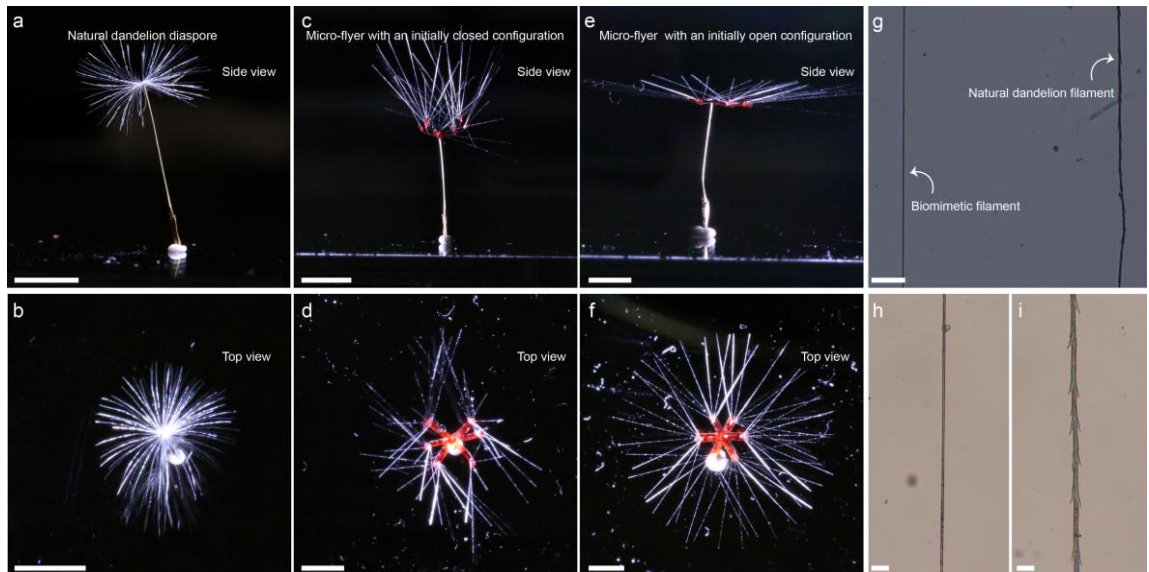

**Supplementary Figure 13. Photographs of the natural dandelion diaspore and the micro-flyers.**
(a) Side-view and (b) top-view of natural dandelion diaspores. (c) Side-view and (d) top-views of the micro-flyer with an initially closed configuration. (e) Side-view and (f) top-view of the micro-flyer with an initially open configuration. All the scale bars in (a-f) are 5 mm. (g) The microscopy images of the biomimetic and natural filaments. The scale bar is 500 μm. The zoomed-in view of the microscopy image of the biomimetic (h) and natural filament (i). The scale bars in (h, i) are 100 μm.

**Supplementary Note 2: Stability of the Separated Vortex Ring.**

The separated vortex ring (SVR) is a toroidal vortex in the wake of the micro-flyer. Its *ad hoc* porous structure allows the SVR to occur for any opening angle and for both symmetric and asymmetric configurations (Supplementary Fig. 14).

The SVR is a stable vortex, *viz.* the flow field forming the SVR has a constant topology at any instant (a toroidal shape with two saddles and two nodes, see Supplementary Fig. 15). If the filamentous structure of the micro-flyer was impervious, the vortex would be unstable, resulting in a von Kármán vortex street (Fig. 2i-2l). Instead, the wake of the micro-flyer shows a stable SVR both at small and large opening angles (Supplementary Fig. 16).

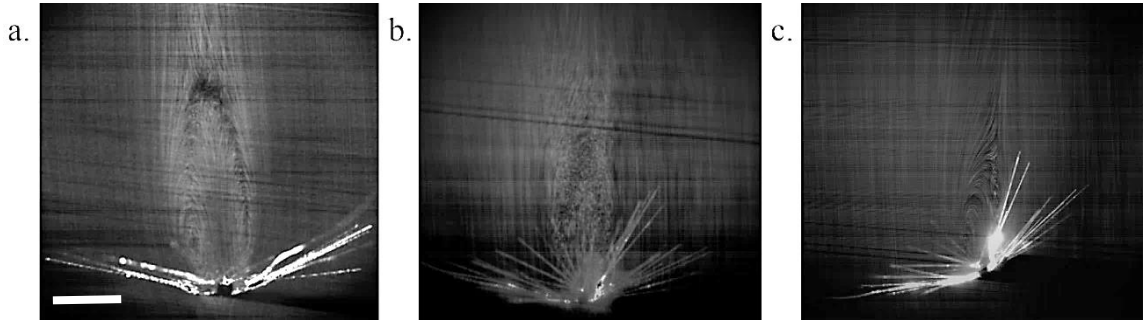

**Supplementary Figure 14. The SVR at different opening angles.** Photographs of the separated vortex ring in the wake of a micro-flyer with an opening angle of 180° (a) and 150° (b), and an asymmetric opening angle of 120° (c). The scale is 5 mm.

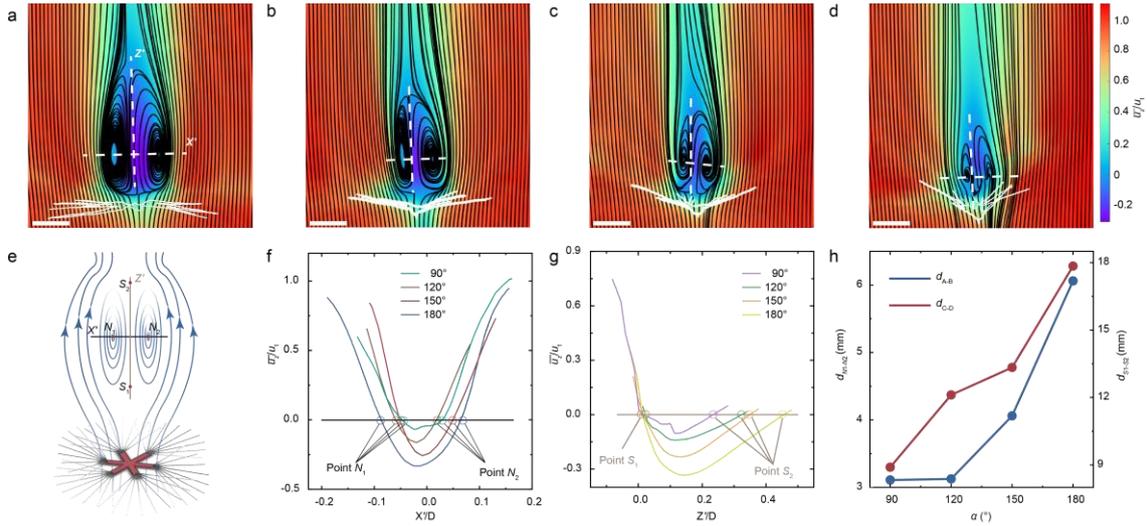

**Supplementary Figure 15. Aerodynamic properties of the vortex ring of the micro-flyer.** Contours of time-averaged vertical velocity on the azimuthal plane of the micro-flyer and streamlines for an opening angle of (a) 180°, (b) 150°, (c) 120°, and (d) 90°. (e) Topological points of the wake, including the left ($N_1$) and right ($N_2$) nodes, and the upstream ($S_1$) and downstream ($S_2$) saddle points. (f) Vertical velocity along the axis $X$ through the nodes $N_1$ and $N_2$ for different values of the opening angle. (g) Vertical velocity along the axis $Z$ through the saddle points $S_1$ and $S_2$ for different values of the opening angle. (h) Euclidean distance between the nodes $N_1$ and $N_2$, and between the saddle points $S_1$ and $S_2$ versus the opening angle. All the scale bars are 0.5 cm.

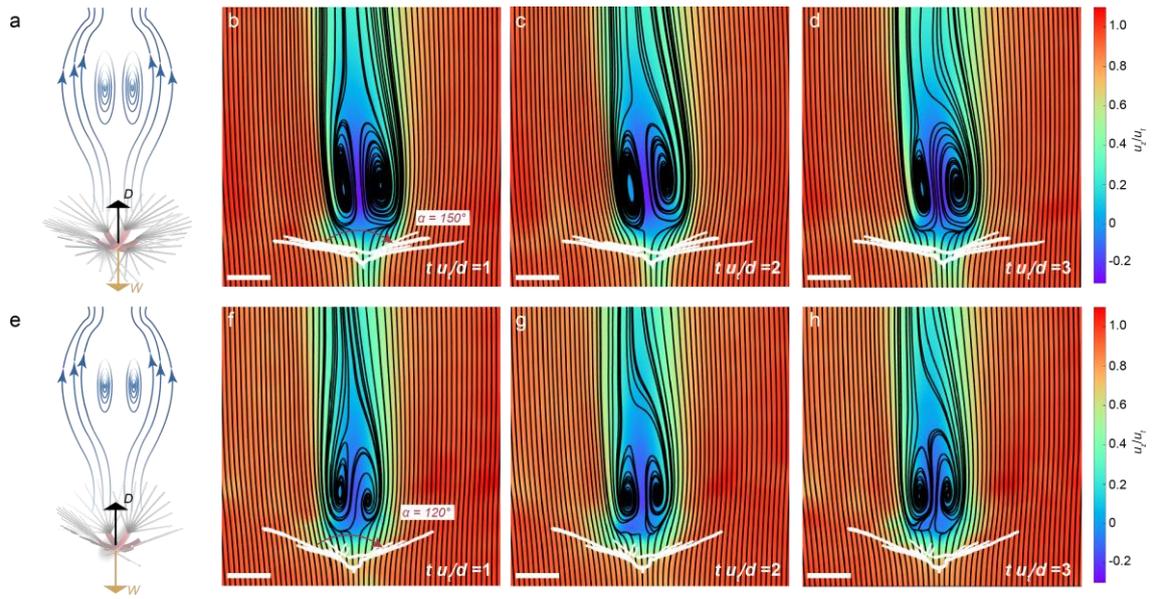

**Supplementary Figure 16. The SVR stability at different opening angles.** The schematic of the micro-flyer with an opening angle of 150° (a). Contours of the instantaneous vertical velocity and streamlines at instants $tu_t/d = 1$ (b), $tu_t/d = 2$ (c), and $tu_t/d = 3$ (d). The schematic of the micro-flyer with an opening angle of 120° (a). Contours of the instantaneous vertical velocity and streamlines at instants $tu_t/d = 1$ (b), $tu_t/d = 2$ (c), and $tu_t/d = 3$ (d). Scale bars are 0.5 cm.

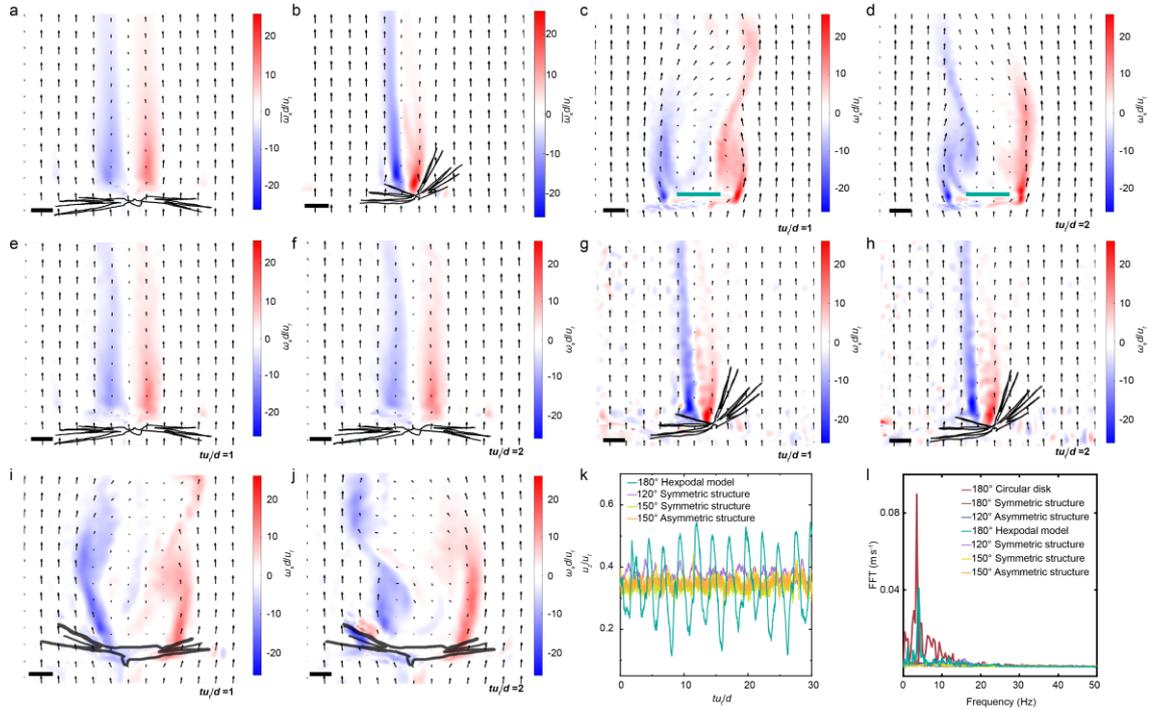

**Supplementary Figure 17. Aerodynamic properties.** (a) Contours of the time-averaged vorticity $\bar{\omega}$ (nondimensionalised with $u_t/d$) around micro-flyers with an opening angle of (a) 180° (fully open) and (b) 120° (asymmetric configuration). Contours of the instantaneous vorticity $\omega$ around the hexapodal core without filaments at two instants one convective period apart, namely (c) $tu_t/d = 1$ and (d) $tu_t/d = 2$. Contours of $\omega$ around a micro-flyer with an opening angle of 180° at (e) $tu_t/d = 1$ and (f) $tu_t/d = 2$. Contours of $\omega$ around a micro-flyer with a 120° asymmetric opening angle at (g) $tu_t/d =1$ and (h) $tu_t/d = 2$. Contours of $\omega$ around a micro-flyer with an added circular disk at its centre at (i) $tu_t/d = 1$ and (j) $tu_t/d = 2$. (k) Time series of the vertical velocity in the wake of various micro-flyers (see legend) measured one diameter downstream ($Z/d = 1$). (l) Fast Fourier transform of the vertical velocity measured in the wake of various micro-flyers ($Z/d = 1$) versus the frequency. Scale bars are 0.5 cm.

**Supplementary Note 3: Aerodynamic Property of micro-flyer.**

The drag coefficient ($C_D$) of the micro-flyer was defined as

$$C_D = \frac{W}{\frac{1}{2}\rho u_t^2 A}, \qquad (1)$$

where the weight $W$ was measured using an Ohaus Explorer analytical balance with a resolution of 0.1 mg. The air density is taken as $\rho = 1.204$ kg m$^{-3}$, corresponding to standard conditions at 20 °C and 1 atm; $A$ is the projected area of the micro-flyer, including both the projected area of the filamentous structure and of the central active component. The terminal velocity $u_t$ was evaluated through drop tests (Method). By adding weight to different models, it was possible to measure how the terminal velocity varied with the weight. This is shown in the Supplementary Fig. 18, where the Reynolds number is

$$Re = \frac{u_t d}{\nu}, \qquad (2)$$

with $d$ being the diameter of the circumscribed circle to the pappus, and $\nu = 15.11 \times 10^{-6}$ m$^2$ s$^{-1}$ is the kinematic viscosity of the air at 20 °C and 1 atm.

To be able to observe the untethered flyer in a fixed position with respect to a laboratory-fixed frame, a vertical wind tunnel was built (Supplementary Fig. 21). By setting the speed of the upward flow stream in the wind tunnel at the same value as the magnitude of the terminal velocity $u_t$, the micro-flyer hovered at a constant position in the tunnel. The wind tunnel velocity was measured by particle image velocimetry.

The micro-flyer was then exposed to different light intensities to change the opening angle (Supplementary Fig. 19). As the opening angle is increased or decreased, the micro-flyer moves upwards or downwards, respectively, because the wind tunnel speed is no longer the same as the terminal velocity. By keeping the illumination constant, and thus the opening angle fixed, the terminal velocity for each opening angle could be measured by identifying the wind tunnel speed allowing the micro-flyer to remain at a constant vertical position in the wind tunnel.

While hovering at a constant height, the micro-flyer rotated around its axisymmetric axis at a modest angular velocity of $1.68 \pm 1.0 \, °\,s^{-1}$ (Supplementary Fig. 23). The rotation behaviour was recorded using a digital single-lens reflex camera (Canon EOS 60D) operating at 25 frames per second. The micro-flyer trajectories were analysed using TRACKER software to determine angular velocity. The remarkable rotatory stability of the micro-flyer is due to the precise axisymmetric structure. In contrast, its natural counterpart, the dandelion seed's diaspore, revealed an angular velocity of $50.8 \pm 17.7 \, °\,s^{-1}$.

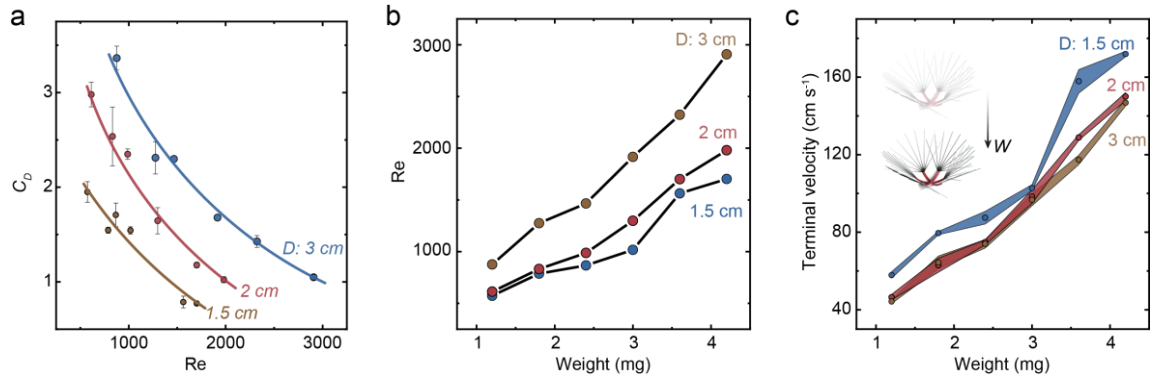

**Supplementary Figure 18. The aerodynamics of micro-flyers.** (a) Drag coefficient of the micro-flyer as a function of the Reynolds number. (b) Reynolds number and (c) terminal velocity of micro-flyers with different diameters under different loads. Error bars are displayed as the mean value +/- one standard deviation ($n = 3$). The same sample was measured repeatedly.

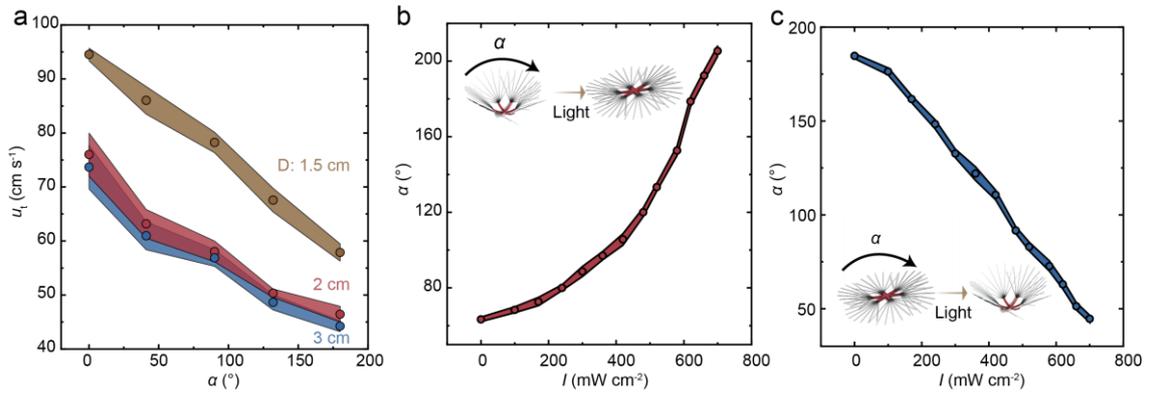

**Supplementary Figure 19. Light-induced shape and terminal velocity changes of micro-flyers.** (a) The variation of terminal velocity at different opening angles for different sample sizes. The change in opening angle in the micro-flyer with an initially (b) closed and (c) open configuration as a function of the light intensity. Error bars are displayed as the mean value +/- one standard deviation ($n = 3$). The same sample was measured repeatedly.

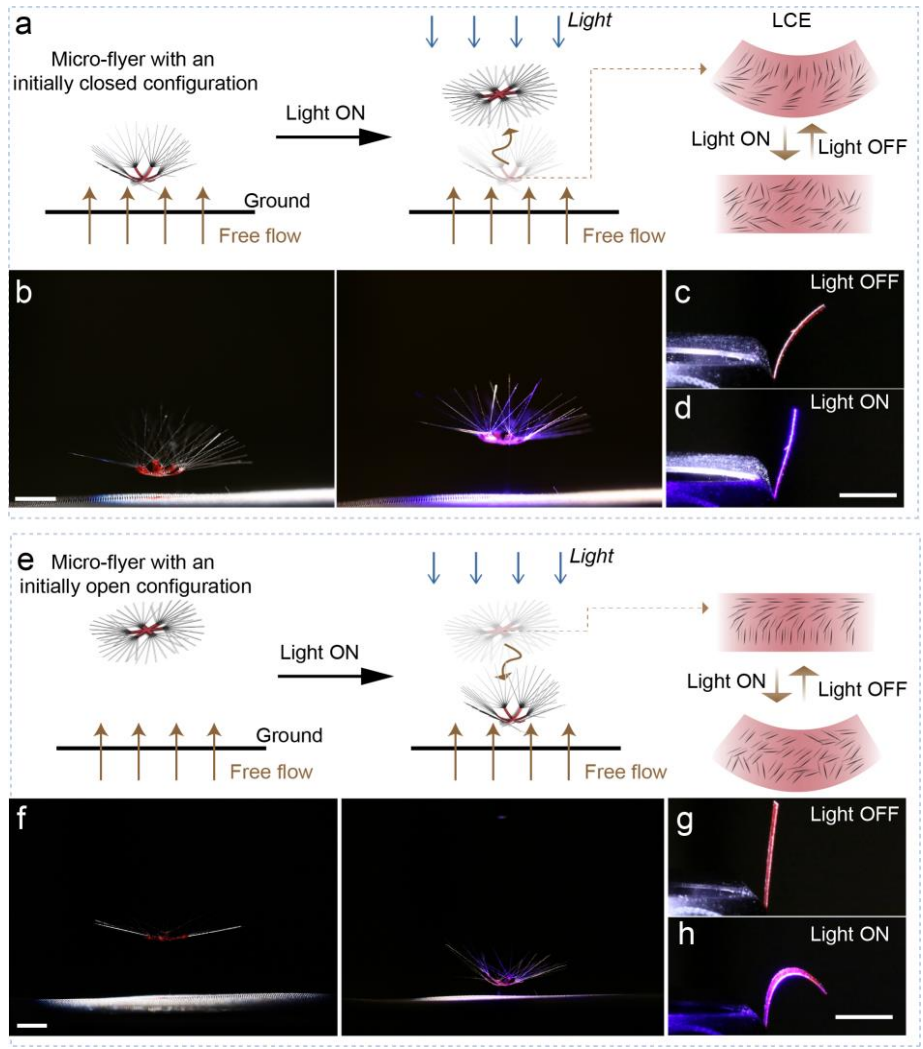

**Supplementary Figure 20. The two distinct structural configurations of the micro-flyer.** (a) The schematic of the micro-flyer with an initially closed configuration, increasing the height under the light illumination. The molecular orientation of corresponding LCE strip with an initially bending configuration is shown on the right. (b) Light-induced upward movement of the micro-flyer with an initially closed configuration. The photographs of (c) a bent LCE strip before light illumination and (d) a flattening strip upon light illumination. (e) The schematic of the micro-flyer with an initially open configuration, decreasing in height under the light. The molecular orientation of the corresponding LCE strip is shown on the right side. (f) Light-induced downward movement of the micro-flyer with an initially open configuration. The photographs of an LCE strip exhibiting (g) a flat shape before light illumination and (h) a bend strip upon light illumination. Wind tunnel speed: 0.6 m s$^{-1}$. The light intensity: 700 mW cm$^{-2}$. All the scale bars are 5 mm.

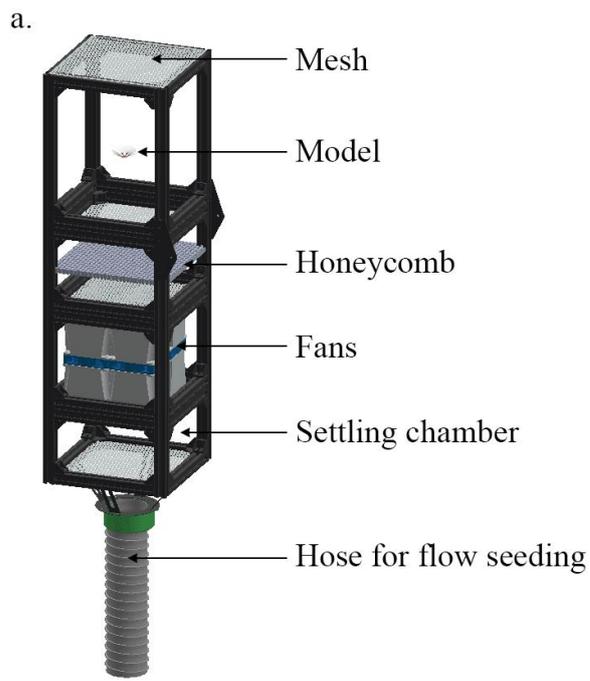 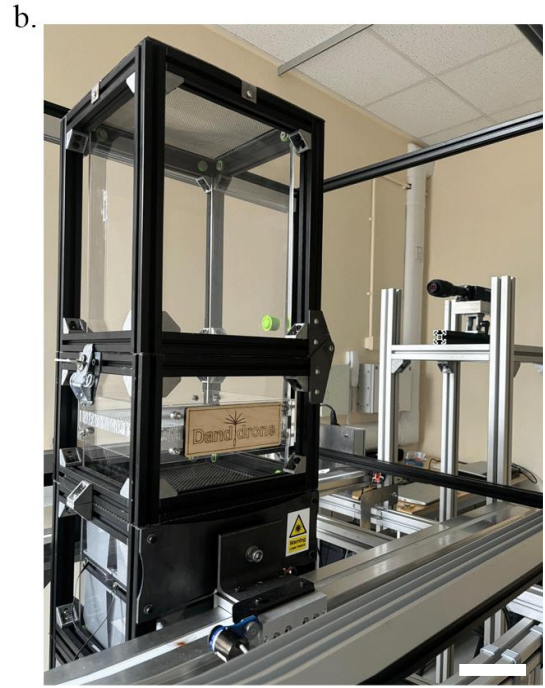

**Supplementary Figure 21. The vertical wind tunnel.** (a) Schematic drawings of the vertical wind tunnel. The model in the test section is not to scale. The air flows in the upward direction. (b) Photograph of experimental setup in laboratory.

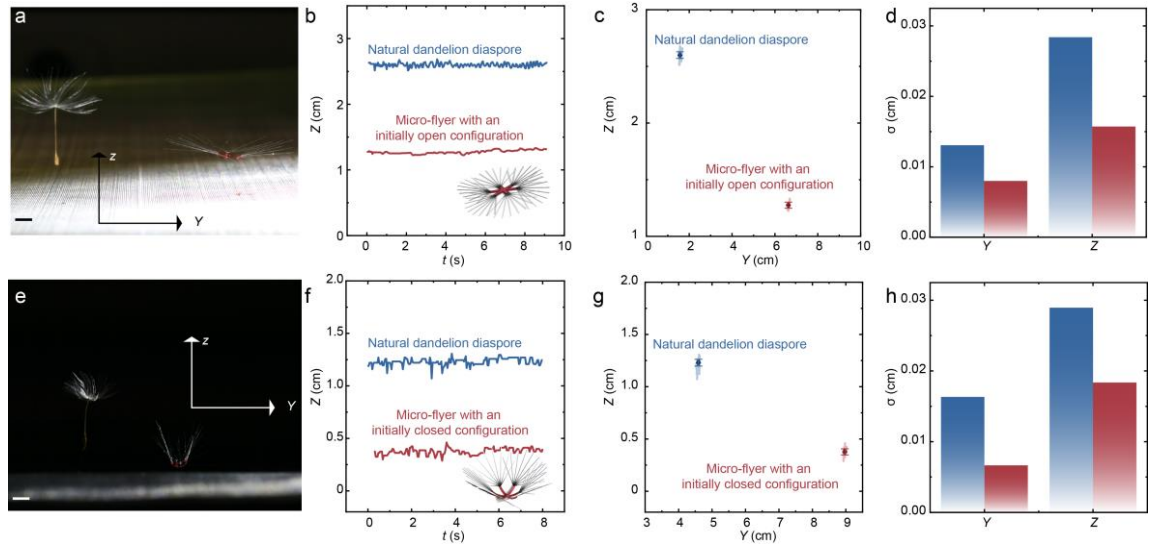

**Supplementary Figure 22. The stability of the micro-flyer.** (a) Photographs of a natural dandelion diaspore (left) and a micro-flyer (right) with an initially open configuration floating in the wind tunnel. (b) The vertical coordinate of the natural dandelion diaspore (blue) and the micro-flyer with an initially open configuration (red) over time. (c) The $Y$ and $Z$ coordinates of the natural dandelion diaspore (blue) and the micro-flyer with an initially open configuration (red) over 1 s (25 data points). (d) The standard deviation of the vertical and horizontal coordinates of the natural dandelion diaspore and the micro-flyer with an initially open configuration. (e) Photographs of a natural dandelion diaspore (left) and a micro-flyer (right) with an initially closed configuration floating in the wind tunnel. (f) The vertical coordinate of the natural dandelion diaspore (blue) and the micro-flyer with an initially closed configuration (red) over time. (g) The $Y$ and $Z$ coordinates of the natural dandelion diaspore (blue) and the micro-flyer with an initially closed configuration (red) over 1 s (25 data points). (h) The standard deviation of the vertical and horizontal coordinates of the natural dandelion diaspore and the micro-flyer with an initially closed configuration. Wind tunnel speed: 0.6 m s$^{-1}$. All the scale bars are 5 mm.

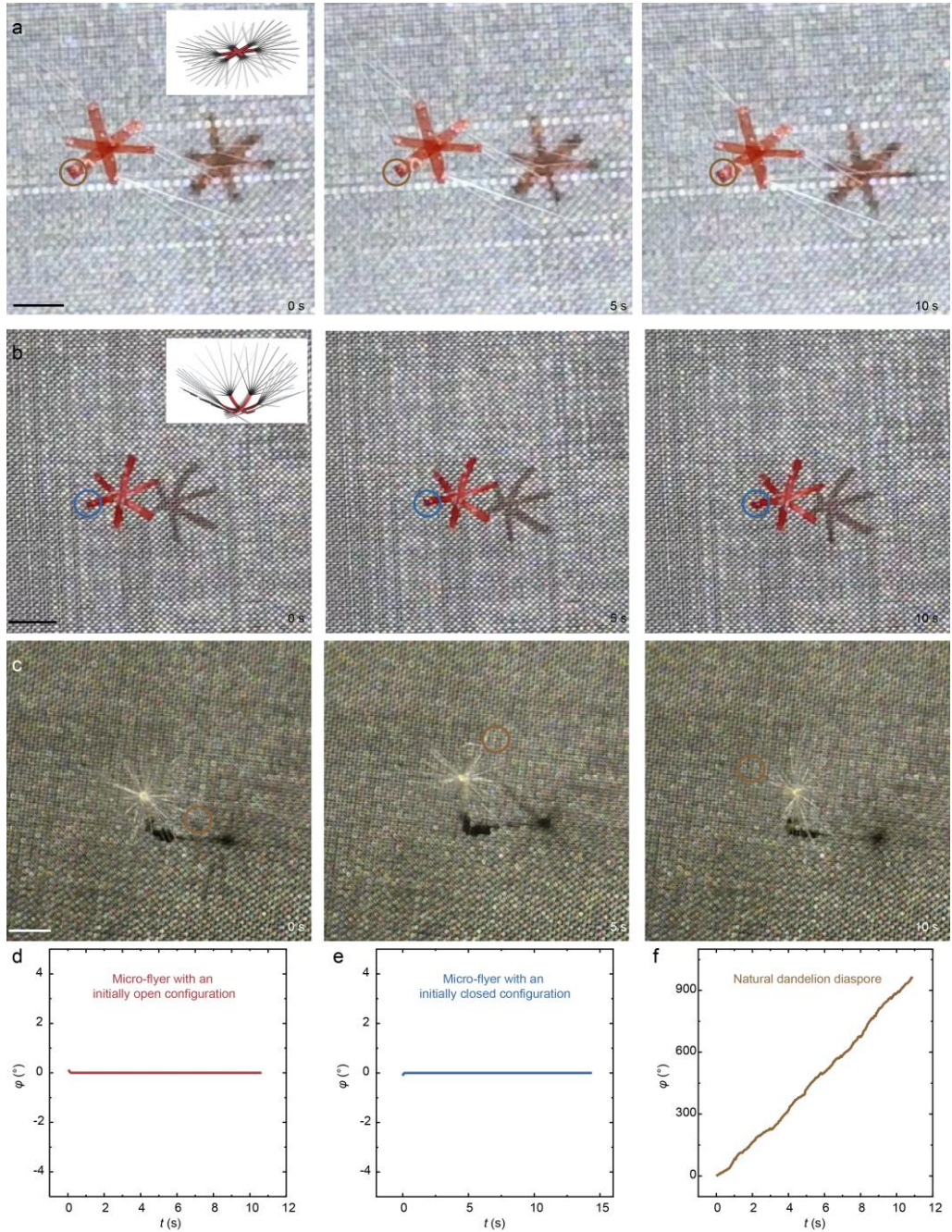

**Supplementary Figure 23. The rotatory stability of micro-flyers.** Snapshots of the micro-flyer with an initially (a) open and (b) closed configuration. (c) Snapshots of a natural dandelion diaspore. The change in rotation angle $\varphi$ of the micro-flyer with an initially (d) open and (e) closed configuration with time. (f) The change of $\varphi$ of a natural dandelion diaspore with time. Wind tunnel speed: 0.6 m s$^{-1}$. Scale bar: 5 mm.

**Supplementary Note 4: The Agile Manoeuvring of the micro-flyer.**

 **Vertical Motion.** Under uniform optical irradiation, the micro-flyer undergoes symmetric deformation (Supplementary Fig. 24), resulting in a change in the aerodynamic drag that enables the vertical motion — either ascent or descent (Supplementary Figs. 25, 26). The direction of motion is determined by the initial geometric configuration (Supplementary Fig. 20). Specifically, for the micro-flyer with an initially open configuration, optical stimulation induces structural closure, leading to a descent motion. In contrast, when the structure begins in a closed state, uniform illumination prompts structural opening, resulting in an ascent motion.

**Flipping.** A uniform sustained illumination on an initially closed state induces progressive opening of the structure, ultimately driving the opening angle to exceed 180° and eventually triggering in-air flipping (Supplementary Fig. 36, 37).

**Target Height.** One can keep a micro-flyer at a constant height by providing illumination at that height (Supplementary Figs. 32, 34). For example, a micro-flyer with an initially closed configuration above the illuminated height falls up to the illuminated height, where it opens and rises again. The micro-flyer exhibits oscillatory behaviour around that altitude, which serves as a tuneable upper or lower bound depending on the initial configuration (Supplementary Figs. 33, 35).

**Horizontal Motion.** In contrast, the selective illumination of different parts of the micro-flyer induces asymmetric deformations (Supplementary Fig. 27), generating lateral propulsion along the horizontal plane (Supplementary Figs. 28, 30). The resulting impulsive translational velocity scales proportionally with light intensity (Supplementary Figs. 29, 31), enabling controllable navigation on the horizontal plane.

Selective local irradiation of specific structural components enables the programming of arbitrary trajectories, such as the formation of letter-like patterns (e.g., A, B, C), as shown in Supplementary Figs. 38-40.

**Three-Dimensional Trajectories.** Together, vertical and horizontal controllable motions allow for the programming of complex three-dimensional trajectories. For instance, a micro-flyer can be elevated to a designated height, translated laterally, and then guided downward (Supplementary Figs. 43, 44), showcasing both the precision and stability of the feedback-driven actuation strategy.

**Clustering.** When two structures are brought into proximity, the negative pressure generated above the micro-flyer leads to mutual attraction, forming a cohesive unit that can be maneuvered as a single entity (Supplementary Fig. 41, 42).

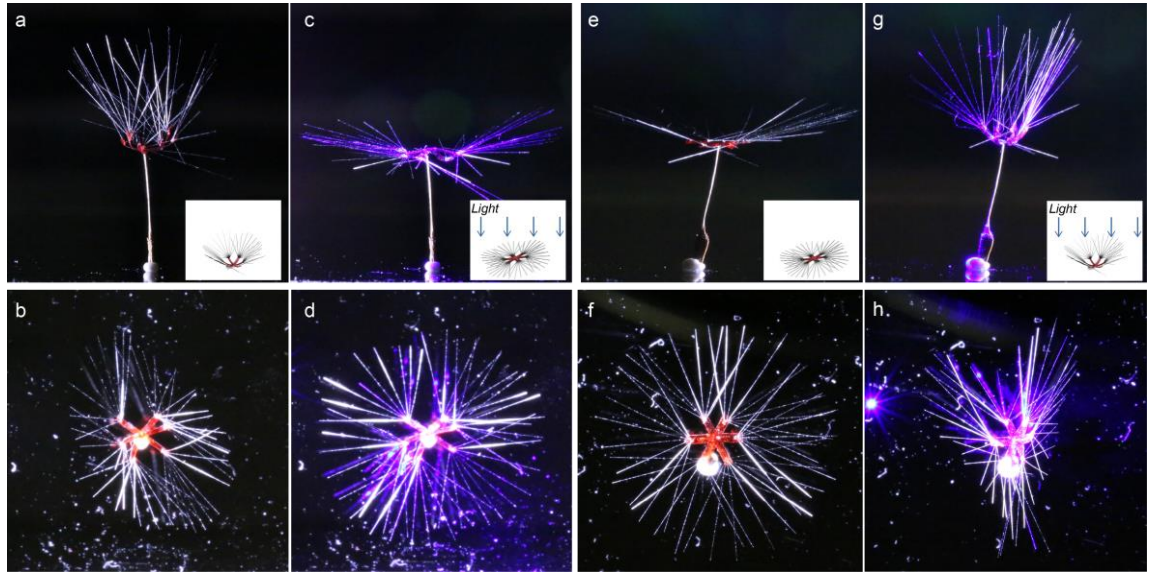

**Supplementary Figure 24. The light-induced symmetric shape-morphing of micro-flyers.** (a) Side-view and (b) top-view photographs of the micro-flyer with an initially closed configuration. (c) Side-view and (d) top-view photographs of the micro-flyer with an initially closed configuration upon light irradiation. (e) Side-view and (f) top-view photographs of the micro-flyer with an initially open configuration. (g) Side-view and (h) top-view photographs of the micro-flyer with an initially open configuration upon light irradiation. Light intensity: 600 mW cm$^{-2}$. Scale bars are 5 mm.

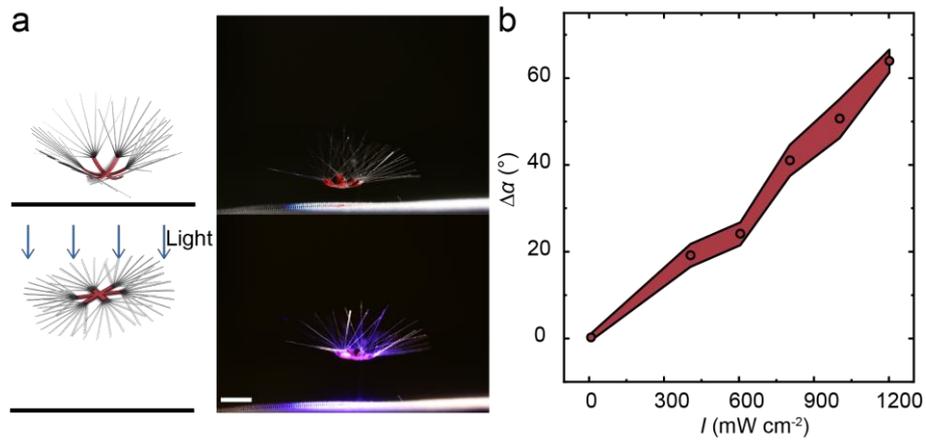

**Supplementary Figure 25. The vertical displacement of an initially closed micro-flyer.** (a) Schematic drawing (left) and pictures (right) of a micro-flyer with an initially closed configuration with light off (top) and on (bottom). (b) The opening angle change of the micro-flyer upon different light intensities. Wind tunnel speed: 0.6 m s$^{-1}$. Error bars are displayed as the mean value +/- one standard deviation (n = 3). The same sample was measured repeatedly.

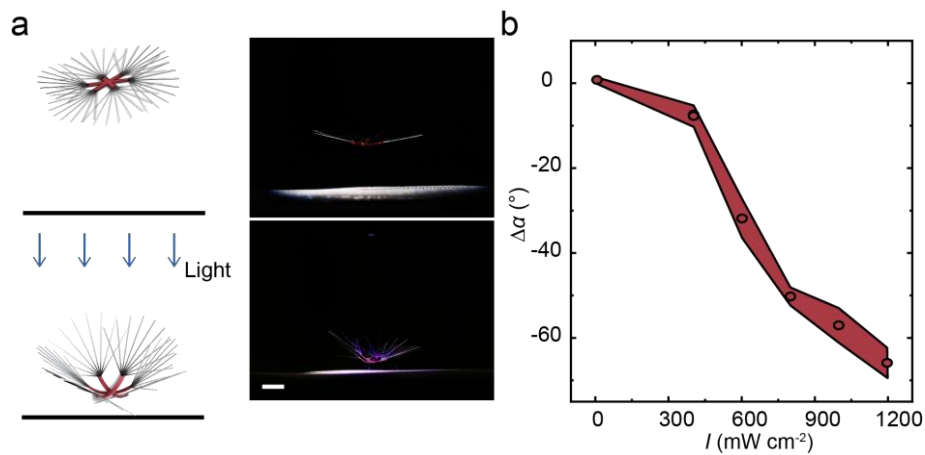

**Supplementary Figure 26. The vertical displacement of an initially open micro-flyer.** (a) Schematic drawing (left) and pictures (right) of a micro-flyer with an initially open configuration with light off (top) and on (bottom). (b) The opening angle of the micro-flyer upon different light intensities. Wind tunnel speed: 0.6 m s$^{-1}$. Error bars are displayed as the mean value +/- one standard deviation (n = 3). The same sample was measured repeatedly.

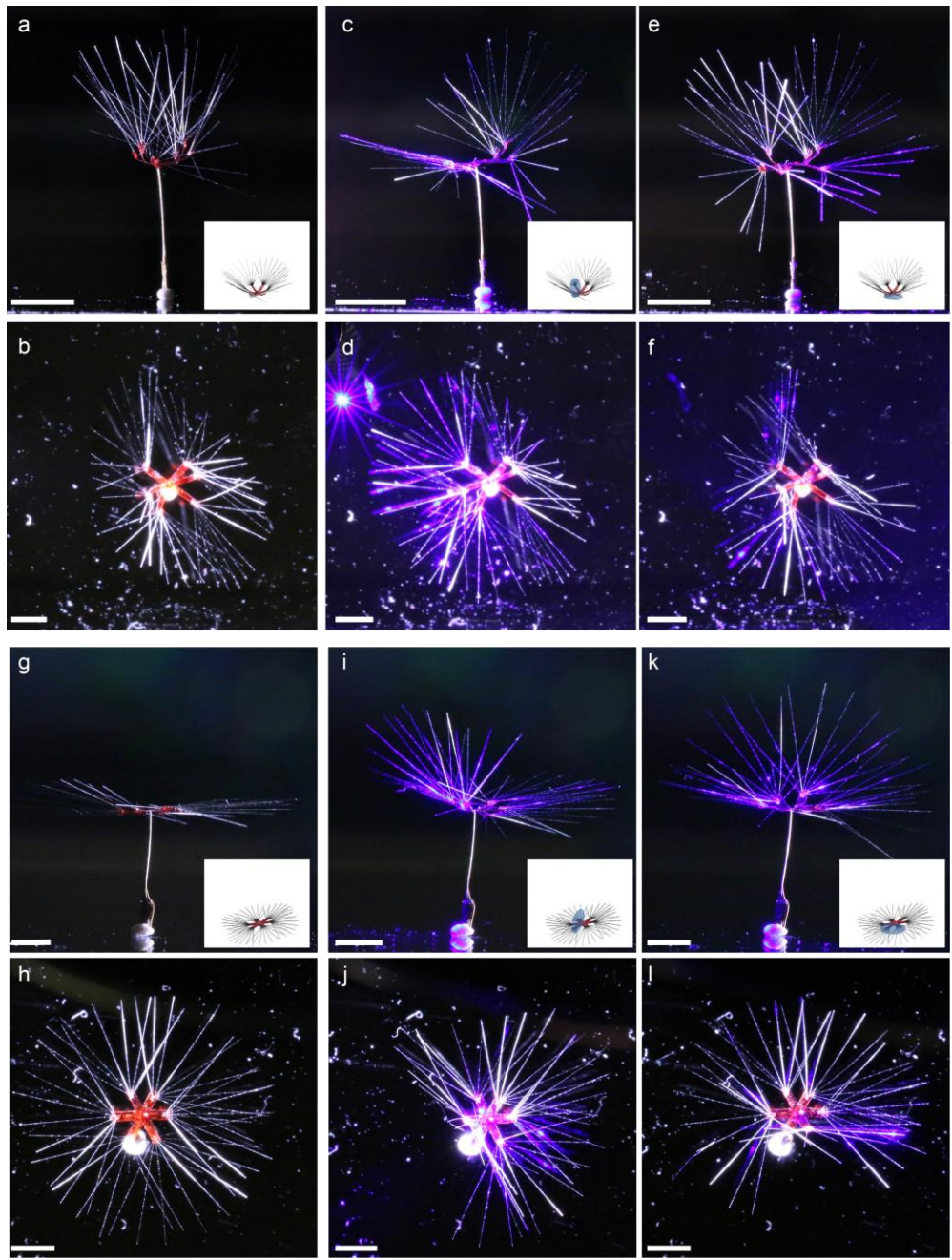

**Supplementary Figure 27. The light-induced asymmetric shape-morphing of micro-flyers.** (a) Side-view and (b) top-view photographs of the micro-flyer with an initially closed configuration. (c) Side-view and (d) top-view photographs of the micro-flyer with an initially closed configuration upon light irradiation on left-side segments. (e) Side-view and (f) top-view photographs of the micro-flyer with an initially closed configuration upon light irradiation on front segments. (g) Side-view and (h) top-view photographs of the micro-flyer with an initially open configuration. (i) Side-view and (j) top-view photographs of the micro-flyer with an initially open configuration upon light irradiation on left-side segments. (k) Side-view and (l) top-view photographs of the micro-flyer with an initially open configuration upon light irradiation on front segments. Light intensity: 600 mW cm$^{-2}$. Scale bars are

5 mm.

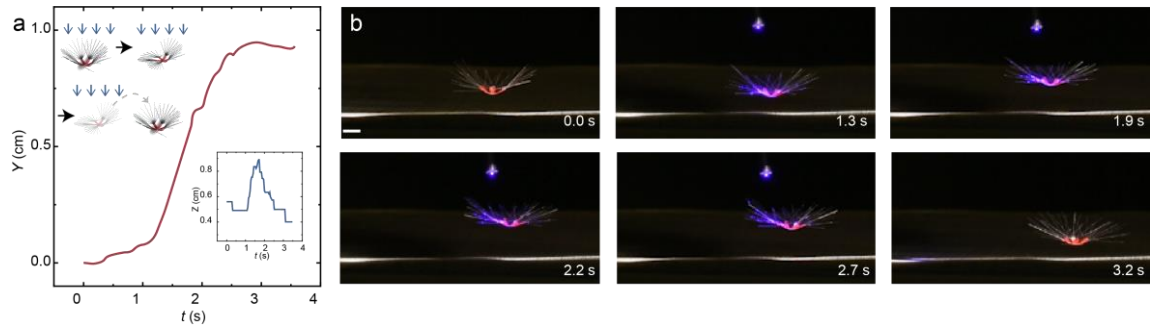

**Supplementary Figure 28. The horizontal displacement of an initially closed micro-flyer.** (a) Time-history of the $Y$ coordinate (and $Z$ coordinate in the inset) of the micro-flyer under light illumination. (b) Snapshot images of the micro-flyer under light illumination. Wind tunnel speed: 0.6 m s$^{-1}$. Light intensity: 600 mW cm$^{-2}$. The scale bar is 5 mm.

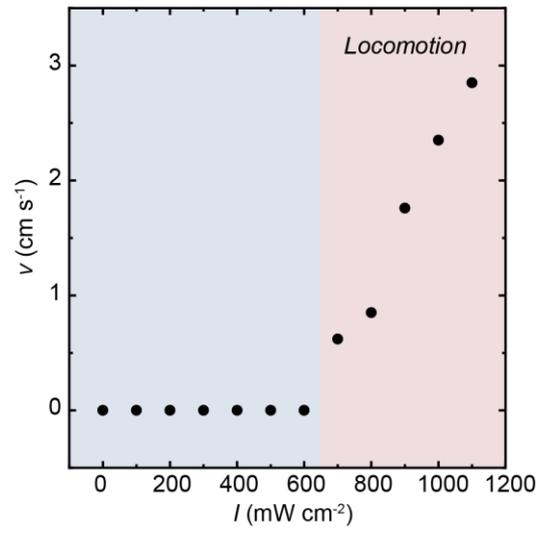

**Supplementary Figure 29. The horizontal speed of an initially closed micro-flyer.** The horizontal speed of the micro-flyer under different light intensities. Wind tunnel speed: 0.6 m s$^{-1}$.

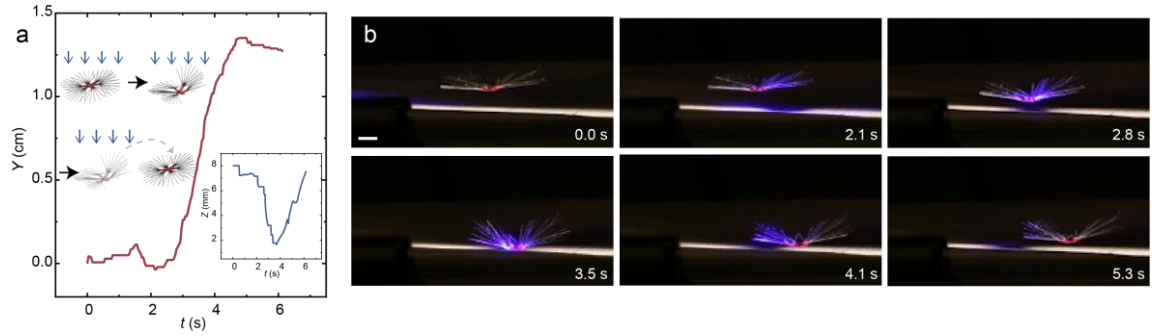

**Supplementary Figure 30. The horizontal displacement of an initially open micro-flyer.** (a) Time-history of the $Y$ coordinate (and $Z$ coordinate in the inset) of the micro-flyer under light illumination. (b) Snapshot images of the micro-flyer under light illumination. Wind tunnel speed: 0.6 m s$^{-1}$. Light intensity: 600 mW cm$^{-2}$. The scale bar is 5 mm.

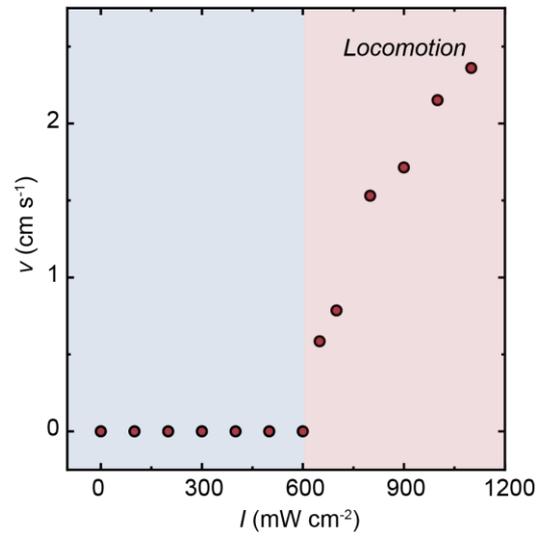

**Supplementary Figure 31. The horizontal speed of an initially open micro-flyer.** The horizontal speed of the micro-flyer under different light intensities. Wind tunnel speed: 0.6 m s$^{-1}$.

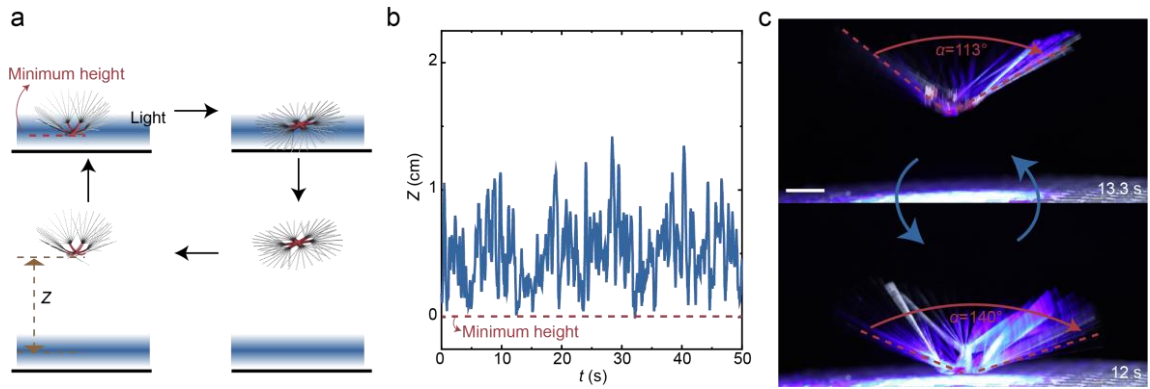

**Supplementary Figure 32. Altitude self-regulation of an initially closed micro-flyer.** (a) Schematic drawing of micro-flyer with an optically set low-altitude limit. (b) Time history of the $Z$ coordinate from the axis of the laser beam. (c) Snapshot images of the micro-flyer's self-regulating opening angle while oscillating around the height limit. Terminal velocity without illumination: 0.8 m s$^{-1}$. Wind tunnel speed: 0.6 m s$^{-1}$. Light intensity: 650 mW cm$^{-2}$. Scale bar is 5 mm.

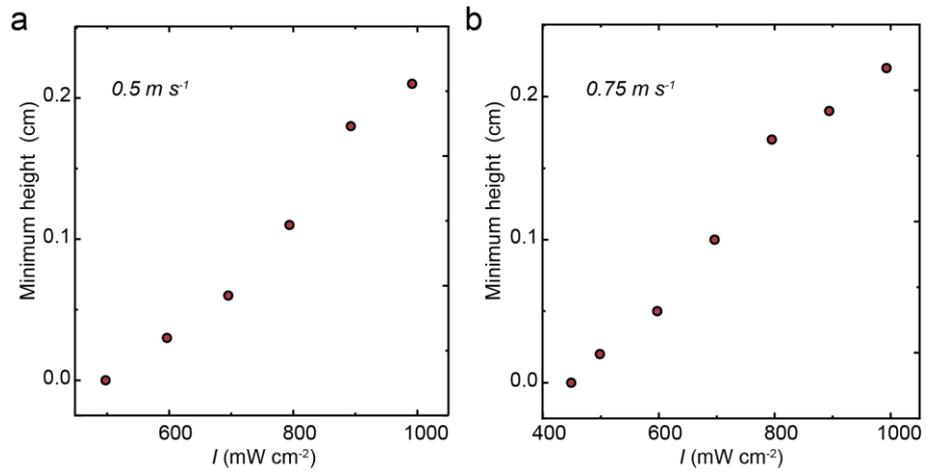

**Supplementary Figure 33. The change of low height limit of an initially closed micro-flyer.** The minimum height of the micro-flyer for different light intensities in a wind tunnel speed of (a) 0.5 m s$^{-1}$, (b) 0.75 m s$^{-1}$ for the same experiment as in Supplementary Figure 32.

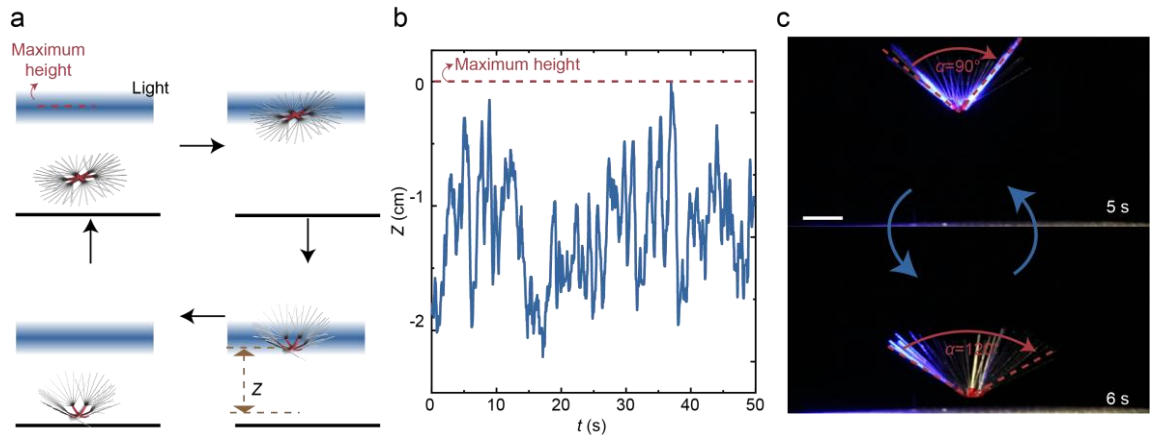

**Supplementary Figure 34. Altitude self-regulation of an initially open micro-flyer.** (a) Schematic drawing of micro-flyer with an optically set high-altitude limit. (b) Time history of the $Z$ coordinate from the axis of the laser beam. (c) Snapshot images of the micro-flyer's self-regulating opening angle while oscillating around the height limit. Terminal velocity without illumination: 0.6 m s$^{-1}$. Wind tunnel speed: 0.75 m s$^{-1}$. Light intensity: 650 mW cm$^{-2}$. Scale bar is 5 mm.

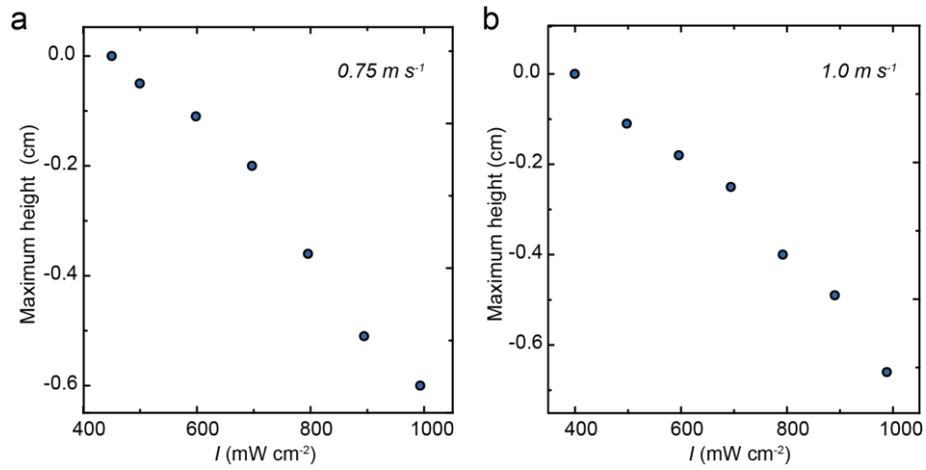

**Supplementary Figure 35. The change of high height limit of the initially open micro-flyer.** The maximum height of the micro-flyer for different light intensities in a wind tunnel speed of (a) 0.75 m s$^{-1}$, (b) 1.0 m s$^{-1}$ for the same experiment as in Supplementary Figure 34.

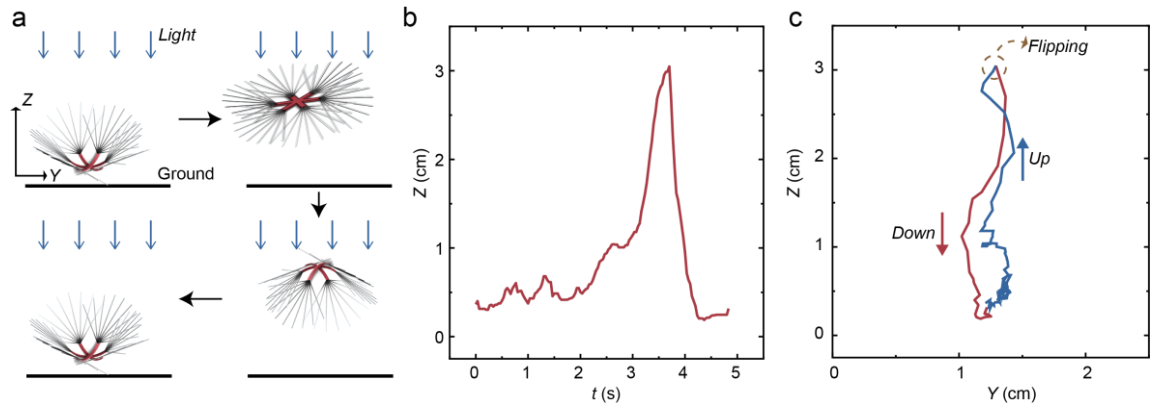

**Supplementary Figure 36. The one-time flipping of a micro-flyer.** (a) The schematic drawing of a micro-flyer flipping in the air. (b) The time history of the *Z* coordinate of the micro-flyer during the flipping process. (c) The trajectory of the micro-flyer flipping in the *Y-Z* plane.

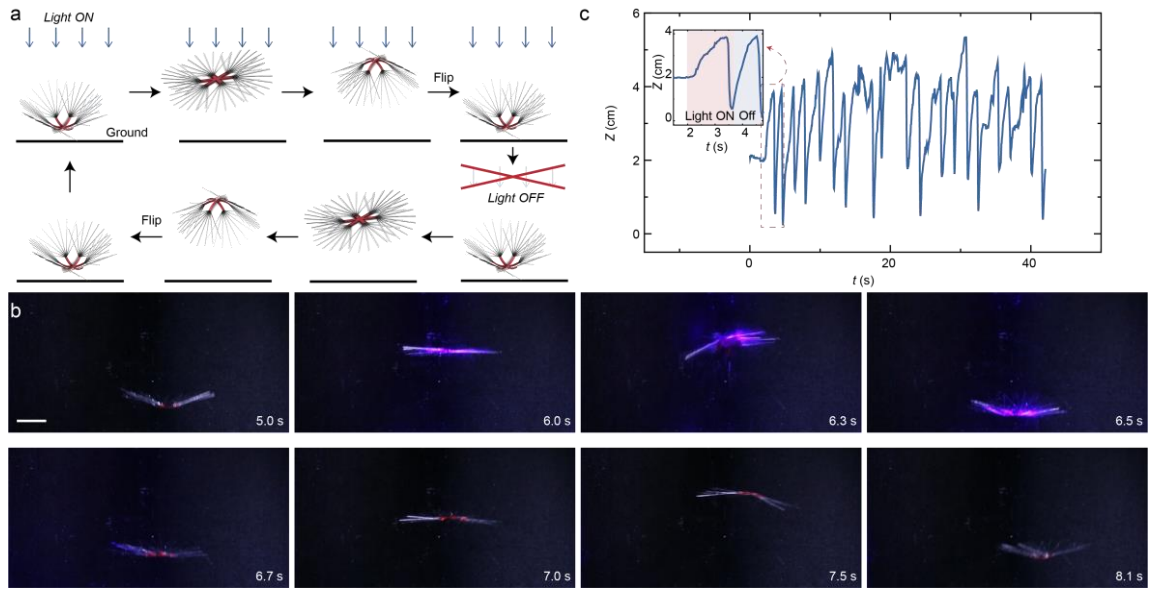

**Supplementary Figure 37. The cyclic body flipping movement of micro-flyer.** (a) The schematic drawing of a micro-flyer flipping upon light illumination and returning to its original position in the dark. (b) Pictures at different times of one cycle of flipping motion driven by light. (c) The $Z$ coordinate of the micro-flyer over more than 40 seconds, including 11 sequential flipping motions. Inset: the zoomed-in view of the micro-flyer's $Z$ coordinate during one flipping cycle. Wind tunnel speed: 0.7 m s$^{-1}$. Light intensity: 800 mW cm$^{-2}$. Scale bar is 5 mm.

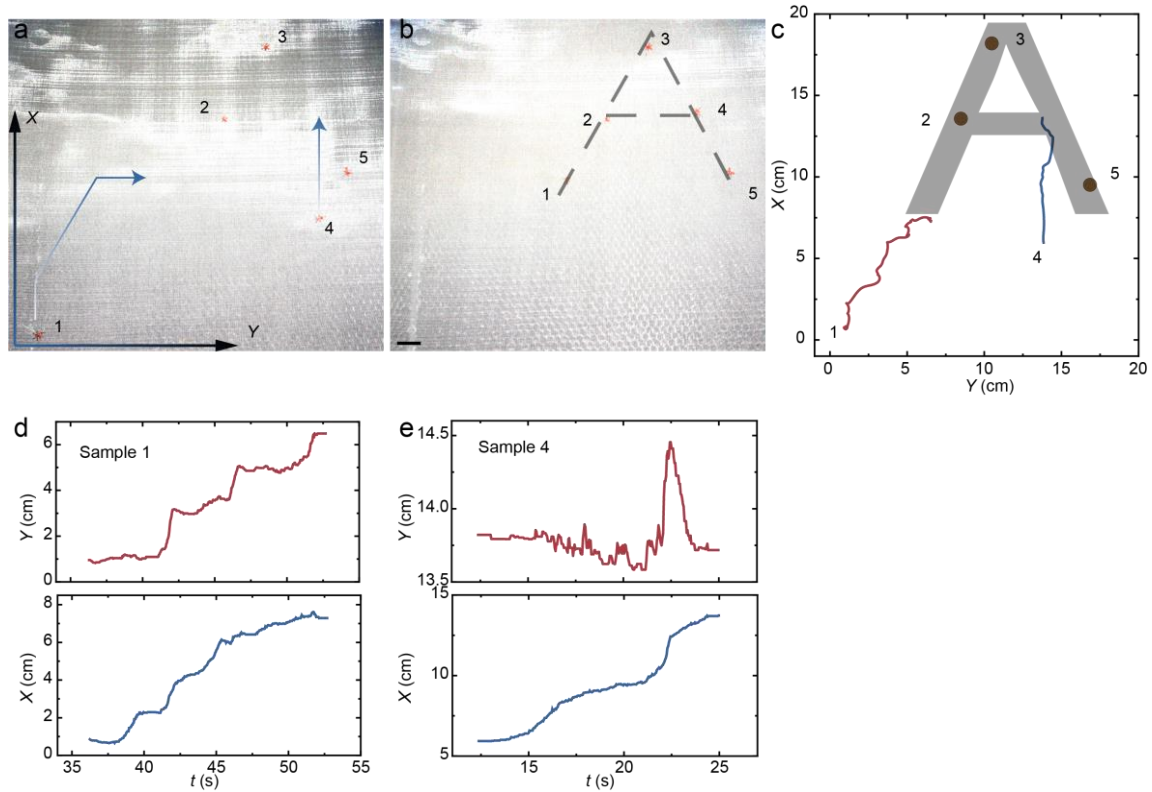

**Supplementary Figure 38. The horizontal locomotion of micro-flyers to form the letter A.** (a, b) Photographs of the micro-flyer moving in the *X-Y* plane. (c) The trajectory of two micro-flyers steered to form the "A" shape. (d) The time histories of the *Y* (top) and *X* (bottom) coordinates of sample 1. (e) The time histories of the *Y* (top) and *X* (bottom) coordinates of sample 4. Wind tunnel speed: 0.6 m s$^{-1}$. Light intensity: 700 mW cm$^{-2}$. Scale bar: 3 cm.

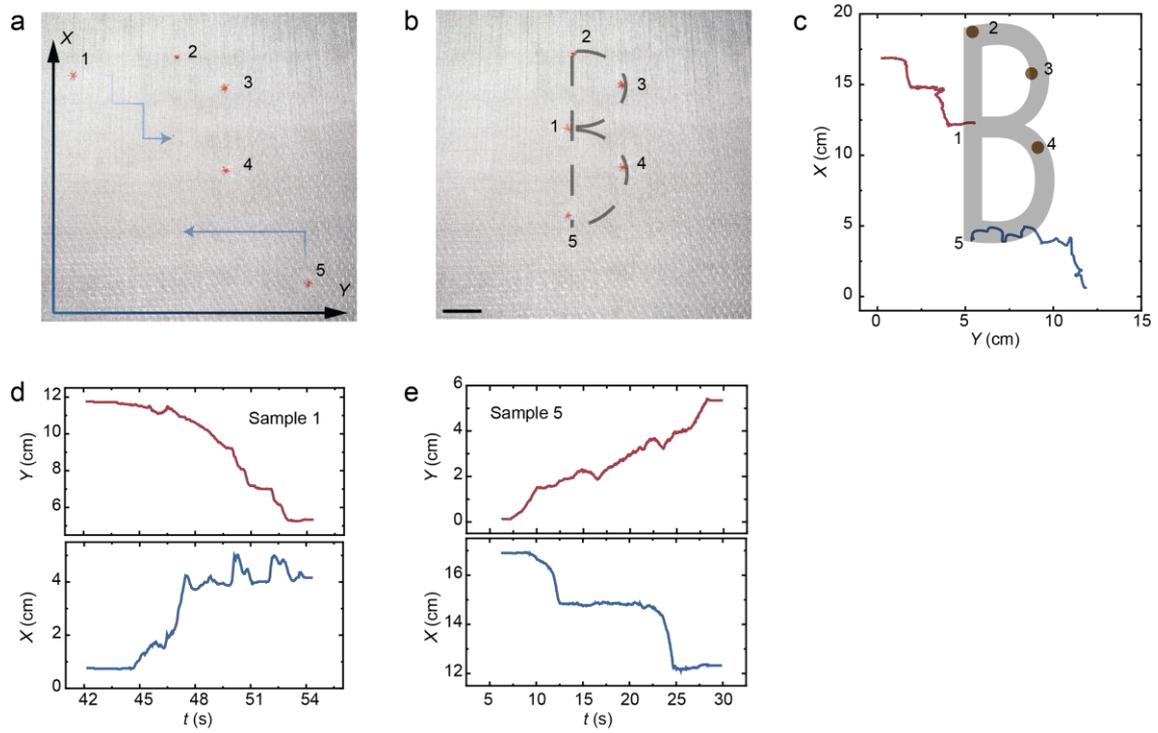

**Supplementary Figure 39. The horizontal locomotion of micro-flyers to form the letter B.** (a, b) Photographs of the micro-flyer moving in the *X-Y* plane. (c) The trajectory of two micro-flyers steered to form the "B" shape. (d) The time histories of the *Y* (top) and *X* (bottom) coordinates of sample 1. (e) The time histories of the *Y* (top) and *X* (bottom) coordinates of sample 5. Wind tunnel speed: 0.6 m s$^{-1}$. Light intensity: 700 mW cm$^{-2}$. Scale bar: 3 cm.

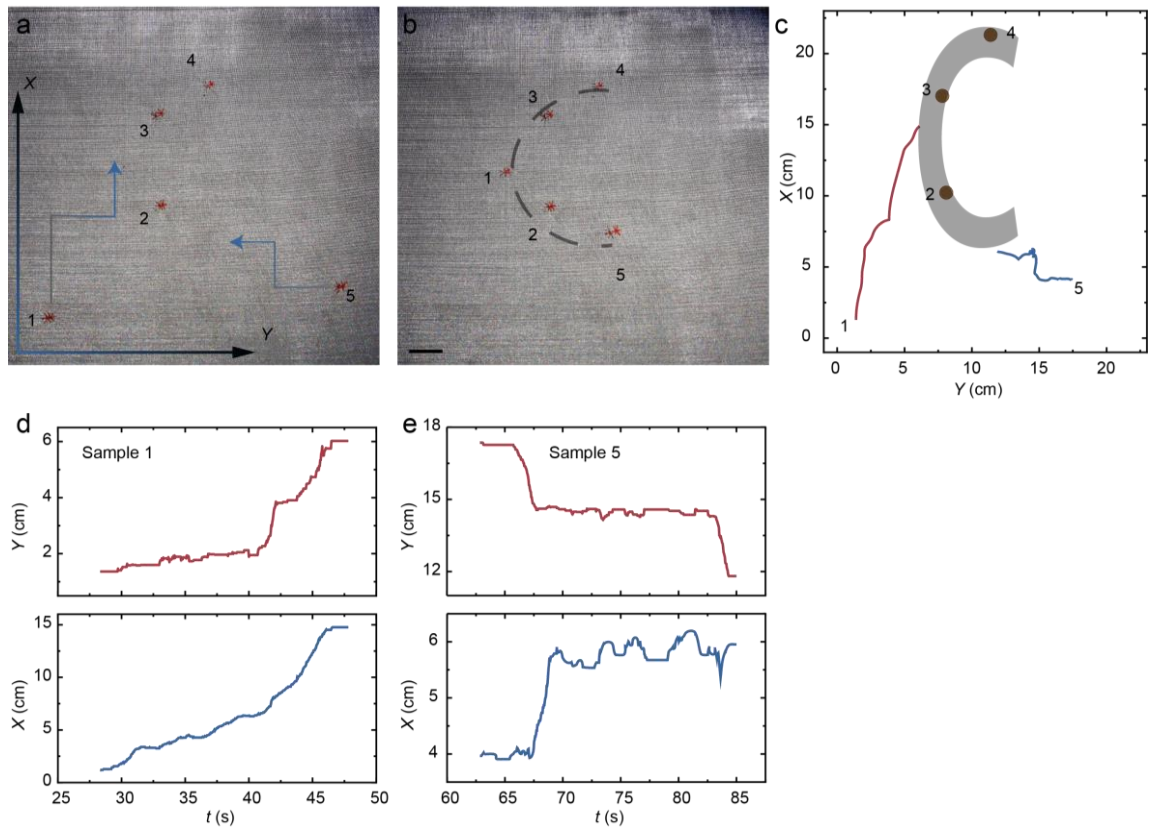

**Supplementary Figure 40. The horizontal locomotion of micro-flyers to form the letter C.** (a, b) Photographs of the micro-flyer moving in the *X-Y* plane. (c) The trajectory of two micro-flyers steered to form the "C" shape. (d) The time histories of the *Y* (top) and *X* (bottom) coordinates of sample 1. (e) The time histories of the *Y* (top) and *X* (bottom) coordinates of sample 5. Wind tunnel speed: 0.6 m s$^{-1}$. Light intensity: 700 mW cm$^{-2}$. Scale bar: 3 cm.

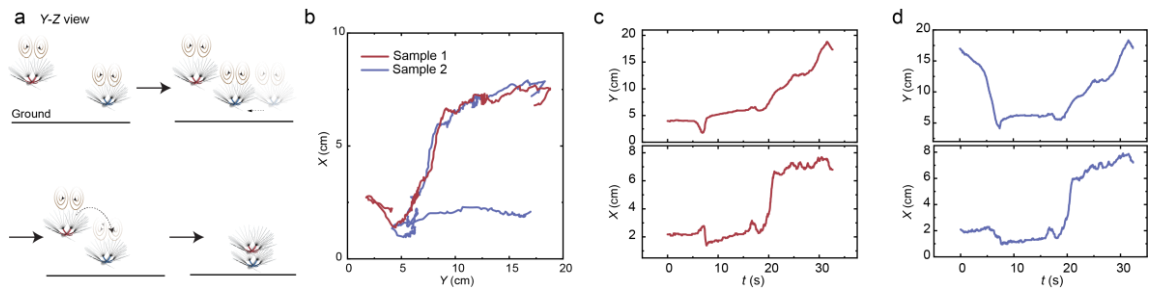

**Supplementary Figure 41. The clustering of two micro-flyers.** (a) The schematic drawing of the clustering between two micro-flyers. (b) The trajectory of two micro-flyers in the *X-Y* plane. (c) The time history of the *Y* (top) and *X* (bottom) coordinates of sample 1. (d) The time history of the *Y* (top) and *X* (bottom) coordinates of sample 2.

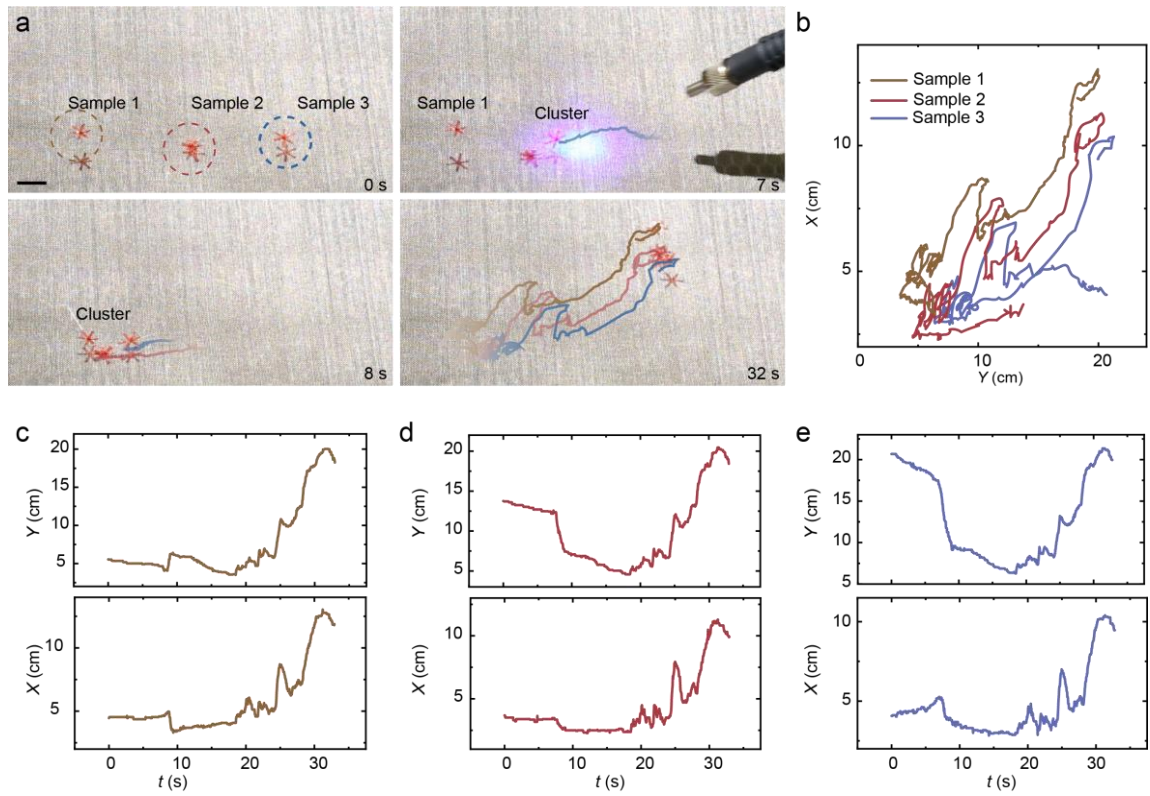

**Supplementary Figure 42. The micro-flyers swarms.** (a) Snapshots of the clustering process among three micro-flyers. (b) The trajectory of three micro-flyers in the *X-Y* plane. The *Y* (top) and *X* (down) coordinates over time for sample 1 (c), sample 2 (d), and sample 3 (e). Wind tunnel speed: 0.6 m s$^{-1}$. Light intensity: 700 mW cm$^{-2}$. The scale bar: 0.5 cm.

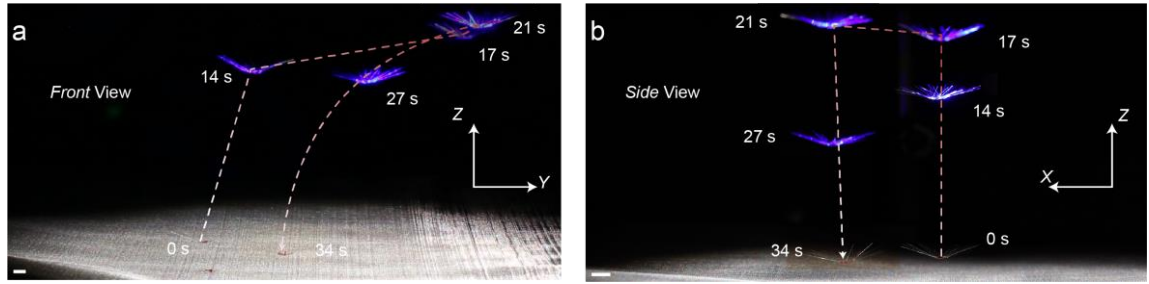

**Supplementary Figure 43. The 3D locomotion of micro-flyer.** Snapshots of three-dimensional of (a) front view and (b) side view of locomotion of micro-flyer driven by light. Wind tunnel speed: 0.6 m s$^{-1}$. Light intensity: 600 mW cm$^{-2}$. Scale bars are 5 mm.

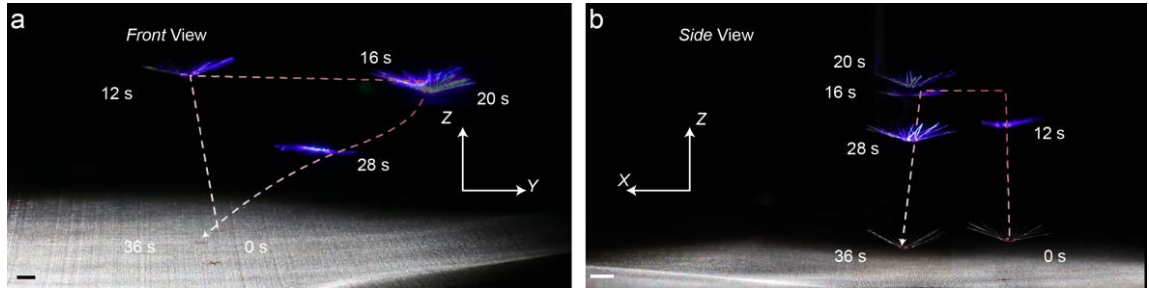

**Supplementary Figure 44. The 3D locomotion of a micro-flyer.** Snapshots of micro-flyers' locomotion through a three-dimensional trajectory; (a) front view and (b) side view. Wind tunnel speed: 0.6 m s$^{-1}$. Light intensity: 650 mW cm$^{-2}$. Scale bars are 5 mm.

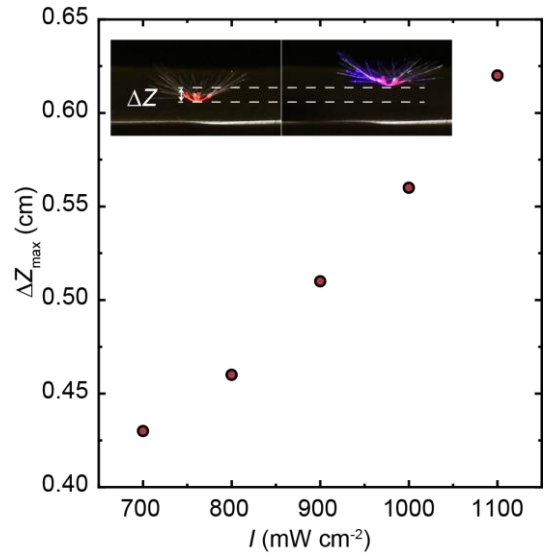

**Supplementary Figure 45. The change in height of an initially closed micro-flyer that is steered to move horizontally by selective illumination of one of its sides.** The change in the $Z$ coordinate of the micro-flyer upon the different light intensities. Wind tunnel speed: 0.6 m s$^{-1}$.

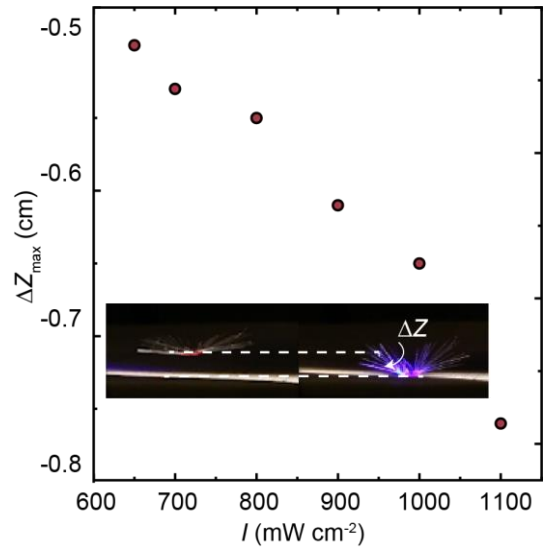

**Supplementary Figure 46. The change in height of an initially open micro-flyer that is steered to move horizontally by selective illumination of one of its sides.** The change in the $Z$ coordinate of the micro-flyer upon the different light intensities. Wind tunnel speed: 0.6 m s$^{-1}$.

**Captions for Supplementary Videos**

**Supplementary Video 1. Light-driven vertical motion of a micro-flyer.**

This real-time video demonstrates the vertical ascent or descent of both micro-flyer configurations under uniform illumination. Light intensity is 700 mW cm$^{-2}$.

**Supplementary Video 2. Light-driven lateral motion of a micro-flyer.**

This real-time video captures the horizontal movement of both micro-flyer configurations under asymmetric illumination. Light intensity is 700 mW cm$^{-2}$.

**Supplementary Video 3. Vorticity contours of a micro-flyer.**

Vorticity contours obtained by PIV at 0.1 × real-time speed (images are acquired at 500 fps and the video is played at 50 fps): (a) fully open (180° opening angle) micro-flyer; (b) asymmetrically oriented micro-flyer with 120° opening angle; (c) fully open micro-flyer with 1 cm disk in the centre; (d) 1 cm disk.

**Supplementary Video 4. Light-induced flipping motion of a micro-flyer in midair.**

This real-time video shows the controlled flipping motion of the micro-flyer upon light stimulation. Light intensity is 700 mW cm$^{-2}$.

**Supplementary Video 5. Light-driven 2D manoeuvring of a micro-flyer.**

In this video, directed light stimuli are used to manoeuvre micro-flyer samples (1 and 4) within a plane, forming the letter "A" in coordination with reference samples (2, 3, and 5). The video is played at 5× speed. Light intensity is 700 mW cm$^{-2}$.

**Supplementary Video 6. Light controlled swarms of micro-flyers.**

This video shows that two or three micro-flyers can be guided to converge, establishing a stable triadic formation that functions as a single, coordinated group for moving. The video is played at 5× speed. Light intensity is 700 mW cm$^{-2}$.

**Supplementary Video 7. Light induced 3D manoeuvring of micro-flyers.**

This video shows the micro-flyer's controlled ascent to a predefined altitude, sustained horizontal navigation, and precise descent back to its original position under light stimulation. The video is played at 5× speed. Light intensity is 700 mW cm$^{-2}$.

79